\documentclass[12pt]{article}
\usepackage[xdvi]{graphicx}
\usepackage{amssymb}
\newcommand{\bea}{\begin{eqnarray}}
\newcommand{\eea}{\end{eqnarray}}

\makeatletter
\@addtoreset{equation}{section}
\makeatother

\newcommand{\simg}{%
\hspace{0.3em}\raisebox{0.4ex}{$>$}\hspace{-0.75em}\raisebox{-.7ex}{$\sim$}\hspace{0.3em}} 

\newcommand{\siml}{%
\hspace{0.3em}\raisebox{0.4ex}{$<$}\hspace{-0.75em}\raisebox{-.7ex}{$\sim$}\hspace{0.3em}}

\def\ignore#1{{}}

\begin{document}
\begin{titlepage}
\begin{flushright}
OU-HET 693/2011
\end{flushright}

\vspace{35ex}

{\Large\bf
~~~Most-attractive-channel study

\vspace{1.5ex}
~~~for bulk and brane fermions in warped space
}

\vspace{5ex}

\begin{center}
{\large
Nobuhiro Uekusa
}
\end{center}
\begin{center}
{\it Department of Physics, 
Osaka University \\
Toyonaka, Osaka 560-0043
Japan} \\
\textit{E-mail}: uekusa@het.phys.sci.osaka-u.ac.jp
\end{center}


\vspace{8ex}

\begin{abstract}

The idea of the self-breaking 
of the standard model gauge symmetry is
applied to a gauge theory in a warped space.
We systematically examine the gauge couplings
of bulk and brane fermions.
The constraint
on the masses of bulk fermions is found 
under the conditions 
that the ordinary four-dimensional massless gluon
is not condensed 
and that color-triplet scalar bound states 
are not formed.
For bulk fermions with zero modes
for the left- and right-handed components, 
the mass parameters are required to be at least
$c\simg 1/2$ and $c\siml -1/2$,
respectively.
Then possible vacuum expectation
values arise only from weak-doublet scalar bound states
which trigger electroweak symmetry breaking.

\end{abstract}

\end{titlepage}


\newpage

\section{Introduction}

In Nature, there are multiple elementary 
particles with
only the values of their masses different from
each other.
It is unknown what the origin of the masses is.
In the standard model of particle physics,
fermions and gauge bosons
acquire their masses
by symmetry breaking through the Higgs mechanism.
While the standard model has passed many tests both 
experimentally and theoretically,
the minimal standard model is valid up to
scales not so high above the weak scale.
In order to describe particle
physics at higher energies,
the standard model needs to be extended.

The direction of the extension
depends on 
whether 
the Higgs boson is fundamental or composite.
If the Higgs boson is fundamental at higher scales,
quantities such as 
the mass squared would significantly run through a
renormalization group flow between the two separate
scales.
Although this effect may drive the mass squared
into a negative value
and various conditions may favor a specific negative value,
 it needs to take into
account that
the running can technically 
pass across zero and other negative values.
If the Higgs boson is composite,
the potential can be suddenly generated.
When
constituents of the Higgs boson are strongly attracted
and form the bound state,
some energy would be released.
The classical background for the Higgs 
boson is stabilized at the minimum of the potential.
The size of the vacuum expectation value
is determined instantly by the condensation
without the above problem of the running.

An economical and interesting idea of the composite
Higgs boson 
was given 
in Ref.~\cite{ArkaniHamed:2000hv}.
The idea is called 
the self-breaking of the standard model
gauge symmetry. 
The point is 
that the gauge bosons of the standard model
propagating in extra dimensions
can rapidly become strongly coupled and
form scalar bound states of quarks and leptons.
The authors proposed that
the existence of a Higgs doublet is a consequence of 
the standard model gauge symmetry and three generation
of quarks and leptons
provided the gauge bosons and fermions
propagate in appropriate extra dimensions compactified
at a TeV scale.
It has also been shown earlier
that electroweak symmetry may be broken
by fields propagating in extra dimensions
\cite{Dobrescu:1998dg}\cite{Cheng:1999bg}.

In this paper, the idea is applied to
a gauge theory in a warped space
whose cutoff is much larger 
than the weak scale or a TeV scale.
Gauge couplings of fermions in the warped space
and phenomenological constraints
have been studied in Ref.~\cite{Davoudiasl:1999tf}
and it has been pointed out that
brane couplings are large compared to
bulk couplings \cite{Chang:1999nh}.
Bound states in the warped space
have been studied in Ref.~\cite{Rius:2001dd}.
Here it has been shown that 
color-triplet composites have 
positive masses squared.
These results above have been given for 
fermions with vanishing bulk masses.
For bulk fermions, 
the values of gauge couplings
are affected by the masses.
The dependence of the bulk gauge couplings
with zero-mode fermions on the bulk mass parameters
has been given in Ref.~\cite{Gherghetta:2000qt}.
If the fourth generation is introduced in the warped
space,
it can be the source of a flavor structure
as well as the dynamical electroweak symmetry
breaking~\cite{Burdman:2007sx}.
Therefore a part of the application of the idea to
the warped space can be seen from the results
in the literature.
The other part requires a new analysis.

We systematically
examine the bulk and brane gauge couplings
of bulk fermions and the brane gauge couplings
of brane fermions.
Our point is that
bulk fermion masses are taken into account
and that
the ordinary four-dimensional massless gluon
is not condensed. 
Unlike Ref.~\cite{ArkaniHamed:2000hv}
with a strict predictive power for the 
top quark mass,
bulk mass parameters and a brane mixing
are included as extra degrees of freedom to 
realize the pattern of quark and lepton masses.
The analysis is performed with the Kaluza-Klein
(KK) mode expansion. 
Binding strengths of fermion constituents
are estimated 
with the most-attractive-channel approximation 
and the evaluation of the gauge couplings.
In particular, 
color-triplet scalar bound states 
composed of a quark and a lepton
need to be avoided.
Once a scenario of the electroweak symmetry breaking
is chosen,
the weak mixing angle, gauge boson masses
and Higgs boson mass are given
in terms of the vacuum expectation values.

The paper is organized as follows.
In Sec.~\ref{sec:mf},
the action integral and mode functions are given.
The boundary values of the mode functions 
are shown explicitly.
In Sec.~\ref{sec:nature},
the field content is given.
The binding strengths
in the most-attractive-channel approximation
are shown.
Numerical evaluation for various bulk
and brane gauge couplings are given in
Sec.~\ref{sec:coupling}.
A scenario for a condensation to trigger
electroweak symmetry breaking
is described.
In Sec.~\ref{sec:bose},
the gauge boson masses,
weak mixing angle and Higgs boson mass
are related to the possible vacuum expectation values.
We conclude in Sec.~\ref{conclusion}.

\section{Action integral
and mode functions \label{sec:mf}}

We consider a gauge theory for a bulk fermion $\Psi$
and a brane fermion $\hat{\Psi}$
in a warped space whose
metric is given by 
$ds^2 = e^{-2\sigma}\eta_{\mu\nu}
     dx^\mu dx^\nu -dy^2$ 
\cite{Randall:1999ee}
\cite{Randall:1999vf}
where
$\sigma= k|y|$ and $|y|\leq L$.
Here $k$ denotes the curvature in 
the five-dimensional anti-de Sitter space.
For the fundamental region $0\leq y \leq L$,
the metric in terms of the coordinate 
$z=e^{ky}$ is written as
$ds^2 =
  z^{-2}
     (\eta_{\mu\nu} dx^\mu dx^\nu - k^{-2}
      dz^2 )$.
Branes are placed at $z=z_0=1$ and 
$z=z_1=e^{kL}\equiv z_L$.
The starting action integral is given by\footnote{%
Even in the flat bulk space,
the five-dimensional Lorentz invariance is violated
at a quantum level for orbifolds
\cite{Uekusa:2010kh}. 
A further generalization in this direction
is left for future work.}
\bea
 I &\!\!\!=\!\!\!& \int d^4x \int_0^{L}
 dy \, \sqrt{\textrm{det}(g_{KL})}
  \, \textrm{tr}
  \left[ - {1\over 2} F_{MN} F_{P Q}
  g^{MP} g^{N Q}
  -{1\over \xi} \omega(A)^2\right]
\nonumber
\\  
  &\!\!\!+\!\!\!&
    \int d^4 x \int_0^{L}
     dy \, \sqrt{\textrm{det}(g_{KL})}
      \, \bar{\Psi} i D \Psi
  +
    u_i \int d^4 x \int_0^{L}
     dy \, \sqrt{-\textrm{det}(g_{\rho\sigma})}
      \, \bar{\Psi} i D \Psi
       \delta(y-y_i) 
\nonumber
\\
  &\!\!\!+\!\!\!&
    \int d^4x \int_0^{L}
      dy \, \sqrt{-\textrm{det}(g_{\rho\sigma})}
       \,
        \bar{\hat{\Psi}}
        i D P_c \hat{\Psi}  \delta(y-y_i) ,
     \label{action}
\eea
where $u_i$ are dimensionful coefficients
and $P_c$ is a chirality projection.
The covariant derivative acts as
\bea
 D \Psi
   &\!\!\!=\!\!\!&
     \left\{
       \Gamma^A
        e_A^M
         (\partial_M + {1\over 8} \omega_{MBC}
            \left[\Gamma^B, \Gamma^C\right]
           - ig_A A_M)
             + i c \sigma' \right\}
               \Psi ,
\\
   D P_c \hat{\Psi}
    &\!\!\!=\!\!\!&
      \Gamma^a e_a^\mu
        (\partial_\mu
          -i g_A A_\mu) P_c \hat{\Psi} ,
    \qquad
          \Gamma^a =
     \left( \begin{array}{cc}
        0 & \sigma^a \\
        \bar{\sigma}^a & 0 \\
        \end{array}\right)
        , \quad
  \Gamma^5  
     = i\gamma^5
\eea
where the spin connection 
and the bulk mass parameter
are denoted as
$\omega_{\mu a 5}
= -\sigma' e^{-\sigma} \delta_{\mu a}$
and $c$, respectively.
Here
$\{ \Gamma^a , \Gamma^b \} = 2\eta^{ab}$,
$\eta = \textrm{diag}(1, -1,-1,-1)$
and $\gamma^5 =\textrm{diag}(-1,1)$.
The gauge fixing functional is given by
$\omega(A) = \partial_\mu A_\nu \cdot
     g^{\mu\nu}
  + \xi z g^{zz} \partial_z (A_z/z)$.
The extension for multiple fields is straightforward.
For the gauge group SU(3) $\times$ SU(2) $\times$ U(1),
the covariant derivative is given by
\bea
   D \Psi = \left\{ \Gamma^A e_A^M
  (\partial_M +{1\over 8} \omega_{MBC} 
 \left[\Gamma^B, \Gamma^C\right]
   -i g_3 G_M -i g_2 W_M
  -i g_1 Y B_M ) 
  + i c \sigma' \right\} \Psi .
\eea 
The gauge fields will be
 denoted as
$A_M = \{ G_M, W_M, B_M\}$, collectively.

In the action integral (\ref{action}),
terms for four fermion interactions
are not included.
Such terms may be induced by
strongly-coupled effects.
Once $\Psi^2$ or $\hat{\Psi}^2$ is regarded as 
a composite scalar $\Phi$ or $\hat{\Phi}$,
the coefficient of $\Phi^2$ or $\hat{\Phi}^2$ term
is fixed by the condensation
without knowing the value of the coefficient
of the original $\Psi^4$ or $\hat{\Psi}^4$ term.
Another aspect of the action integral is that
the bulk fields have 
the bulk and brane terms.
In addition to bulk fermions, brane fermions
contributions to 
brane terms.
Both of the gauge couplings for the bulk
fermion $\Psi$ and the brane fermion $\hat{\Psi}$
are flavor diagonal. 
A part of fields $\Psi$ and $\hat{\Psi}$
can become heavy and the other can become light,
depending on the quantum numbers.
This gives rise to a flavor mixing
\cite{Burdman:2002se}\cite{Uekusa:2008iz}. 
Furthermore, the coefficient of
the brane kinetic term is not an ambiguous parameter.
This is contrastive to the coefficients of 
irrelevant operators which are 
ambiguous and unfixed within the present framework.
For example, one of ways to achieve the coefficient
is to employ the property of chirality.
The on-shell renormalization requires  
a propagator to evenly decompose for two chiralities
while corrections on orbifolds differ between
left- and right-handed fermions. 
Hence, the two independent renormalization coefficients
are required.
In the flat space, the wave function renormalization
and the coefficient of the brane kinetic term are
shown to be
unambiguously fixed \cite{Uekusa:2010jf}.
For simplicity,
the brane kinetic term and
the $\Psi$-$\hat{\Psi}$ mixing will be omitted
in the present analysis
whereas brane gauge couplings will be taken into
account for both of $\Psi$ and $\hat{\Psi}$.

We examine the case where the only possible source of
gauge symmetry breaking 
is condensation.
The boundary conditions are imposed 
as Neumann for the 4-component
$A_\mu$ and Dirichlet for the extra-dimensional
component $A_y$. 
The boundary conditions for fermions will be assigned 
depending on their 
quantum numbers.
Identical boundary conditions are adopted
at the positions of the two branes.
For example,
a fermion with the Neumann condition 
for the left-handed component
at $y=0$
obeys the Neumann condition for $y=L$.
Then the left-handed component has a zero mode.
In the $\xi=1$ gauge, the gauge field is
decomposed into
\bea
    A_\mu (x,z) 
    &\!\!\! =\!\!\!& {1\over \sqrt{L}} 
  \left[ A_{\mu 0} (x)
       + \sum_{n=1}^\infty N_n A_{\mu n} (x) 
  \chi_n (z)  \right],
\\
    A_z (x,z) 
    &\!\!\! =\!\!\!& {1\over k\sqrt{L}} 
  \sum_{n=1}^\infty N_n A_{z n} (x) 
  \phi_n (z)  .
\eea
The mode functions are given by
\bea
   \chi_n = z F_{1,0} (m_n z/k, m_n z_L/k)
   ,
\qquad
    \phi_n = z F_{0,0} (m_n z/k, m_n z_L/k)
    .
\eea
Here the function $F$ is defined as
$F_{\alpha,\beta} (u,v)
 \equiv
 J_{\alpha}(u) Y_\beta (v) -Y_\alpha (u) J_\beta(v)$ 
for non-negative $u,v$.
It satisfies
$F_{-n,-m}(u,v) = (-1)^{n+m} F_{n,m}(u,v)$
for $n,m=0,1,2,\cdots$, 
and
\[
   F_{-\gamma, -\delta} (u,v)
    = \sin [(\delta-\gamma)\pi]\cdot
       \left[ J_\gamma (u) J_\delta(v)
         + Y_\gamma (u) Y_\delta (v)\right] 
   + \cos [(\delta-\gamma)\pi]
     \cdot F_{\gamma, \delta} (u,v) ,
\]
where $\gamma, \delta$ are not integers.
Hence,
$F_{-\alpha, -\alpha}(u,v) = F_{\alpha,\alpha}(u,v)$.
The $n$-th KK gauge boson mass 
is obtained from the eigenvalue equation
\bea
   F_{0,0} (m_n/k, m_n z_L/k)=0 .
    \label{gmasseq}
\eea
The normalization is given by
$N_n^{-2} 
 =  \int_1^{z_L} dz \,  \chi_n^2/ (k L z)$.

Except for the $\delta$ function part,
the equations of motion for fermions are given by
\bea
      i \sigma
           \cdot \partial \tilde{\Psi}_R
        + k 
        D_+\tilde{\Psi}_L        = 0 ,
    \qquad
      i \bar{\sigma} \cdot
       \partial \tilde{\Psi}_L
     -k D_- \tilde{\Psi}_R
     = 0 ,
\eea 
where $\tilde{\Psi} \equiv z^{-2} k^{-1/2}\Psi$.
The bulk fermion is decomposed into 
\bea
  \tilde{\Psi} (x,z)
   = f_0 (z) \psi_0 (x) +
  \sum_{n=1}^\infty  N_{f n} f_n (z) \psi_n(x) .
\eea 
The normalization is given by
$N_{fn}^{-2} = \int_1^{z_L}
   dz \, f_n^2$.
The mode functions obey
\bea
       (D_+ D_- - m_n^2/k^2) f_{Rn} = 0 ,
 \qquad
     (D_- D_+ - m_n^2/k^2) f_{Ln}  = 0
   .
\eea  
Here 
$D_ + \equiv z^c \partial_z z^{-c} \cdot$
and
$D_- \equiv -z^{-c} \partial_z z^c \cdot$.
The $\delta$ function part depends on
the boundary conditions at $z=z_i$.

\subsubsection*{R-even}

When the right-handed component obeys 
the Neumann boundary condition,
we call its fermion an R-even fermion and
the $\delta$ function part means
$D_- \tilde{\Psi}_R \bigg| = 0$
at the boundary.
The mode functions are given by
\bea
 \textrm{even} &&
   f_{Rn}
    = \sqrt{z} F_{c+(1/2), c-(1/2)}
      (m_n z/k, m_n z_L/k) ,
\\
  \textrm{odd} &&
  f_{Ln} = 
  \sqrt{z} F_{c-(1/2), c-(1/2)}
     (m_n z/k, m_n z_L/k) .
\eea
The left-handed component has 
the Dirichlet boundary condition.
The zero mode exists for an even component,
$f_{R 0} =
  [(1-2 c)/(z_L^{1-2 c} - 1)]^{1/2}
  \,  z^{-c}$.
The eigenvalue equation for the $n$-th KK fermion is
given by
\bea
  F_{c-(1/2), c-(1/2)}
   (m_n /k, m_n z_L/k) = 0.
    \label{rmasseq}
\eea
The mass for $(c-1/2)$ is 
the same as the mass for $-(c-1/2)$.
For example, the mass for $c=2$ 
is the same as the mass for $c=-1$.
The eigenvalue equation for a KK gauge boson,
(\ref{gmasseq})
is the same as 
for an R-even fermion with $c=1/2$.

\subsubsection*{L-even}
When the left-handed component obeys 
the Neumann boundary condition,
we call its fermion an L-even fermion and
the $\delta$ function part means
$D_+ \tilde{\Psi}_L \bigg| = 0$
at the boundary.
The mode functions 
are given by
\bea
  \textrm{even} &&
    f_{L n} =
      \sqrt{z} F_{c-(1/2), c+(1/2)}
       (m_n z/k, \lambda m_n z_L/k) ,
\\
  \textrm{odd} &&
     f_{R n} = \sqrt{z}
      F_{c+(1/2), c+(1/2)}
        (m_n z/k, m_n z_L/k) .
\eea
The right-handed fermion has 
the Dirichlet boundary condition.
The zero mode exists for an even component,
$f_{L 0} =
    [(1+ 2 c)
      / (z_L^{1 + 2 c} -1)]^{1/2}
      \, z^c$.
The eigenvalue equation for the $n$-th KK fermion is
given by
\bea
   F_{c+(1/2), c+(1/2)}
    (m_n /k, m_n z_L/k) = 0 .
\eea
From the property of $F$,
this equation equals $F_{-c-(1/2), -c-(1/2)}
    (m_n /k, m_n z_L/k) = 0$,
which is 
Eq.~(\ref{rmasseq}) with $c\leftrightarrow -c$.
Therefore, the mass for an L-even fermion
with $c$
is equal to the mass for an R-even fermion with $-c$.

KK masses are calculated numerically.
One input is the warp factor.
The assumption is 
that the five-dimensional geometry is 
a classical background.
We regard the cutoff at $y=0$ as
an intermediate scale which is much lower than
the Planck scale or the stabilization of geometry
as in Ref.~\cite{Uekusa:2010ge}.
The warp factor is adopted as $z_L=10^{10}$.
In order that the KK scale 
$m_{\textrm{\scriptsize KK}}
=\pi k/(z_L -1)$ is of the order 
${\cal O}(1)$TeV, as the other input
the curvature is taken as
$k = 4 \times 10^{12}$.
Then KK masses are given 
with the parameter $c$ in Table~\ref{tab:kkmass}.

\begin{table}[htb]
\begin{center}
\caption{KK masses in unit of GeV.
The fermion masses are shown for R-even fermions.
\label{tab:kkmass}}
\vspace{1ex}
\begin{tabular}{c|ccccc|c}
\hline\hline
KK mode &
1 & 2 & 3 & 4 & 5 
& $c$ \\ \hline
Gauge boson &
9896.51 & 
22372 & 
34913.4 & 
47469.3 &
60030.4 &
(1/2) \\
\hline
Fermion &
23053.8 & 
36380 &
49291.8 & 
62058.4 &
74756.1 &
$-2$
\\
&
17973.6 &
30901 &
43616.5 &
56264.8 & 
68883 &
$-1$ 
\\
& 
12566.4 & 
25132.7 &
37699.1 &
50265.5 &
62831.9 &
0 
\\
\hline
\end{tabular}
\end{center}
\end{table}

\begin{figure}[htb]
\begin{center}
\includegraphics[width=7.5cm]{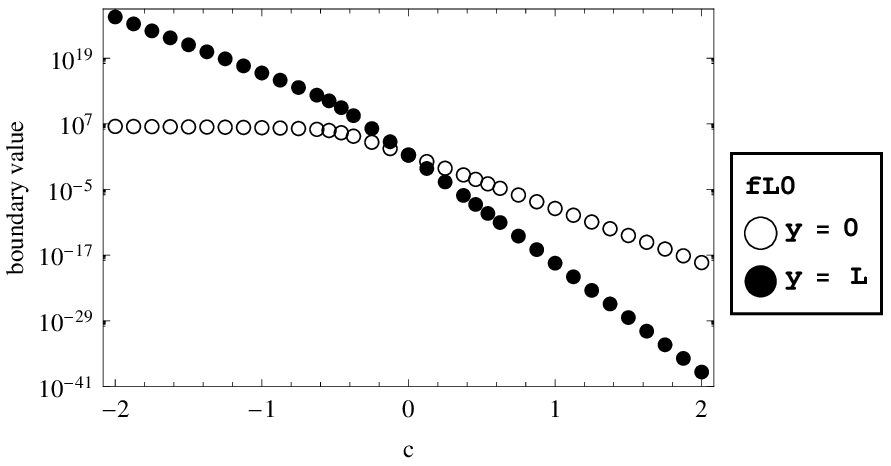} \qquad
\includegraphics[width=7.5cm]{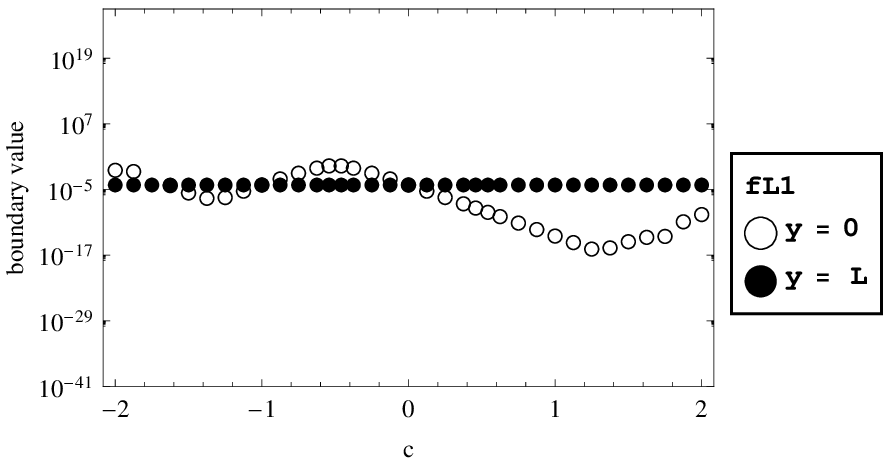}

\includegraphics[width=7.5cm]{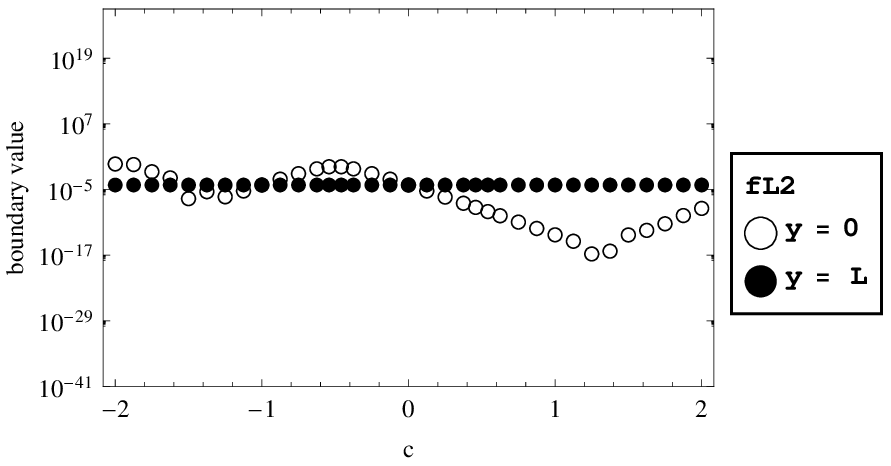} \qquad
\includegraphics[width=7.5cm]{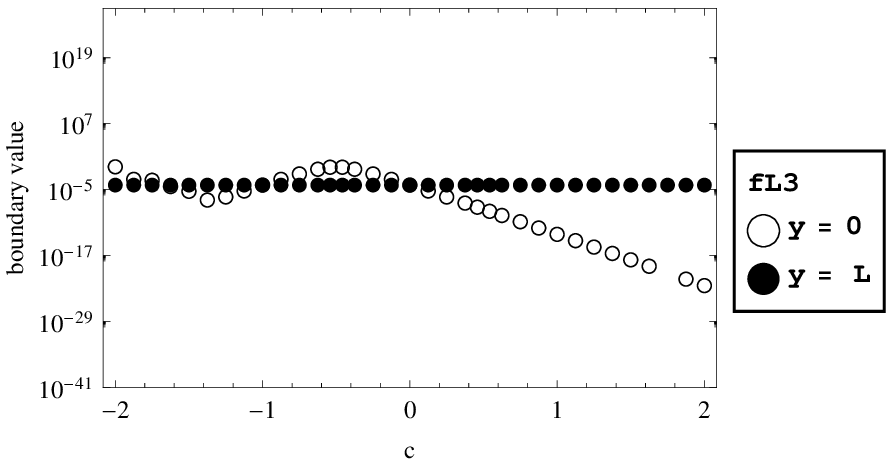}
\caption{Boundary value $f_{Ln}|_{y=0,L}$
where $n=0,1,2,3$.
\label{fig:flbv}}
\end{center}
\end{figure}

For the same input values,
the boundary values for the mode functions 
are shown in Table~\ref{tab:chi}
and Fig.~\ref{fig:flbv}.
\begin{table}[hb]
\begin{center}
\caption{Boundary value $N_n \chi_n|_{y=0,L}$.
\label{tab:chi}}
\begin{tabular}{c|ccccc}
\hline \hline
KK mode &
1 & 2 & 3 & 4 & 5 \\
\hline
$y=0$ &
$-0.24066$ &
0.167176 &
$-0.136784$ &
0.119079 &
$-0.107114$ 
\\
$y=L$ &
6.79041 &
6.7882 &
6.78752 &
6.78719 &
6.78699
\\
\hline
\end{tabular}
\end{center}
\end{table}
The mode functions for the gauge boson
have small values for $y=0$
and large values for $y=L$.
If there are gauge couplings on the brane with
a TeV-scale cutoff,
their values tend to receive 
large contributions from the gauge boson part.

The boundary values of mode functions for fermions
depend on
whether they are KK modes or zero mode.
When they are KK modes,
the dependence on the KK level is not large.
For $y=L$, the boundary values are small
and almost independent of $c$.
For $y=0$, they are also small in a wide region.
On the other hand,
zero mode significantly depends on $c$.
For a positive and large $c$,
the boundary value is small.
The mode function for a R-even fermion, $f_R$
is given by 
$f_L$ with the replacement $c\to -c$.
For a positive and large $c$,
the boundary value for R-even $f_{R0}$ is large.

In the next section,
we will consider 
composite objects with the form $\bar{\Psi}_1 \Psi_2$.
If the constituent $\Psi_1$ with a mass parameter
$c_1$ is an L-even fermion and
$\Psi_2$ with $c_2$ is an R-even fermion,
$\bar{\Psi}_1 \Psi_2$ has zero mode.
After the condensation, the composite scalar 
can give rise to a potential with
a negative mass squared
as pointed out 
in Refs.~\cite{ArkaniHamed:2000hv}\cite{Rius:2001dd}.
The $y$-dependent 
profile of the zero mode for $\bar{\Psi}_1 \Psi_2$
is given by $\bar{\Psi}_{1L0} \Psi_{2R0} \sim 
z_L^{-(c_1-c_2)y/L}$.
The localized position for the zero mode contributions
of the composite 
is determined by the difference 
$c_1-c_2$.
For $c_1> c_2$,
they are localized at the brane with 
an intermediate-scale cutoff.
It is  an open question whether the condensation 
localized at the brane with an TeV-scale cutoff
significantly affects low energy physics.
However, a scenario to respect
the standard model at low energies 
is a high-energy condensation
and it may be natural to expect that it corresponds
to $c_1 \geq c_2$.
If all we have to do is 
to reproduce the pattern of observed masses,
it might be fulfilled even for $c_1 = c_2$
for five-dimensional Dirac fermions
corresponding to 
weak-doublet and singlet fields such as
$e_L$ and $e_R$
\cite{Uekusa:2009pj}.

\section{Matter content and attractive force \label{sec:nature}}

Our standpoint is to systematically 
examine attractive forces for various possible 
fermions rather than to build a model with 
a fixed set of fermions.
The field contents are similar to the standard model
based on SU(3) $\times$ SU(2) $\times$ U(1).
One generation is treated.
Instead of the four-dimensional chiral fermions,
fields are
bulk Dirac fermions and brane chiral fermions.
The fields and their quantum numbers are
 given in Table~\ref{tab:matter}.
For the bulk fields, each zero mode is shown. 
\begin{table}[htb]
\begin{center}
\caption{Bulk and brane matter \label{tab:matter}}
\vspace{.5ex}
\begin{tabular}{lll}
\hline \hline
5D bulk field (zero mode) & Quantum number & 4D brane field \\
\hline
$Q_{0} = \left(\begin{array}{c}
t \\ b \\ \end{array}\right)_L$
& $(3,2,{1\over 6})$ 
& $\hat{Q} = \left(\begin{array}{c}
\hat{t} \\ \hat{b} \\ \end{array}\right)_L$
\\
$U_{0} = t_R$ 
& $(3,1,{2\over 3})$
& $\hat{U}=\hat{t}_R$ 
\\
$D_{0} = b_R$
& $(3,1,-{1\over 3})$
& $\hat{D}=\hat{b}_R$
\\
$L_{0} = \left(\begin{array}{c}
\nu_\tau \\ \tau \\ \end{array}\right)_L$
& $(1,2,-{1\over 2})$
& $\hat{L} = \left(\begin{array}{c}
\hat{\nu}_\tau \\ \hat{\tau}\\ \end{array}\right)_L$
\\
$N_{0} =\nu_{\tau R}$
& $(1,1,0)$
& $\hat{N}=\hat{\nu}_{\tau R}$
\\
$E_{0} = \tau_R$
& $(1,1,-1)$
& $\hat{E} =\hat{\tau}_R$
\\
\hline
\end{tabular}
\end{center}
\end{table}

The SU(2)-doublet bulk fermions $Q$, $L$ are 
L-even fermions
and the singlet bulk fermions $U$, $D$, $N$, $E$ are
R-even fermions.
Each Dirac fermion has the masses
$c_f$ where $f=Q,L,U,D,N,E$.
These parameters generate the hierarchical 
pattern of the observed masses.
Following Ref.~\cite{ArkaniHamed:2000hv},
we perform the most-attractive-channel 
approximation.
The coefficient of the potential for the
attractive force for $\bar{\Psi}_1\Psi_2$
is given by \cite{Raby:1979my}
\bea
   {1\over 2} \left[
    C_2 (\bar{\Psi}_1)
      +C_2 (\Psi_2)
      -C_2 (\bar{\Psi}_1\Psi_2) \right] ,
      \label{c2}
\eea
where $C_2(r)$ is the second Casimir invariant for
the representation $r$ of the gauge group.
Possible combinations for 
$\bar{\Psi}_1\Psi_2$ for bulk fermions
are shown in Table~\ref{tab:bs}.
\begin{table}[htb]
\begin{center}
\caption{Binding strength for $\bar{\Psi}_1\Psi_2$.
\label{tab:bs}}
\begin{tabular}{c|cccc}
\hline\hline
 & Constituents
& SU(3)$\times$ SU(2)$\times$U(1)
& Binding
& Relative binding
\\
&
& representation
& strength
& for $\sqrt{3\over 5}\,g_1 = g_2=g_3$ 
\\
\hline
$H_1$ & $\bar{Q}U$ 
& $(1,2,{1\over 2})$
& ${4\over 3} g_3^2 + {1\over 9} g_1^2$
& 1 
\\
$H_2$ & $\bar{Q}D$
& $(1,2,-{1\over 2})$
& ${4\over 3} g_3^2 -{1\over 18} g_1^2$
& 0.93 
\\ \hline\hline
$S_1$ & $\bar{Q}Q$
& $(1,1,0)$ 
& ${4\over 3} g_3^2 + {3\over 4}g_2^2 
+ {1\over 32} g_1^2$ & 1.5
\\
$S_2$ & $\bar{U}U$
& $(1,1,0)$
& ${4\over 3} g_3^2 + {4\over 9} g_1^2$
& 1.14
\\
$S_3$ & $\bar{D}D$
& $(1,1,0)$
& ${4\over 3} g_3^2 + {1\over 9} g_1^2$
& 1
\\
\hline 
$S_4$ & $\bar{U}D$
& $(1,1,-1)$
& ${4\over 3}g_3^2 - {2\over 9} g_1^2$
& 0.86
\\ \hline
$S_5$ & $\bar{L}L$
& $(1,1,0)$
& ${3\over 4}g_2^2 + {1\over 4} g_1^2$
& 0.64
\\
$S_6$ & $\bar{E}E$
& $(1,1,0)$ 
& $g_1^2$
& 0.21
\\ \hline
$S_7$ & $\bar{L}Q$
& $(3,1,{2\over 3})$
& ${3\over 4}g_2^2 -{1\over 12}g_1^2$
& 0.5
\\
$S_8$ & $\bar{E}U$
& $(3,1,{5\over 3})$
& $-{2\over 3} g_1^2$
& $-0.29$ 
\\
$S_9$ & $\bar{E}D$
& $(3,1,{2\over 3})$
& ${1\over 3}g_1^2$
& 0.14
\\
\hline \hline
$S_{10}$ & $\bar{E}Q$
& $(3,2,{7\over 6})$
& $-{1\over 6}g_1^2$
& $-0.07$
\\
$S_{11}$ & $\bar{L}U$
& $(3,2,{5\over 6})$
& 0 & 0
\\
$S_{12}$ & $\bar{L}D$
& $(3,2,{1\over 6})$
& ${1\over 6}g_1^2$
& 0.07
\\
$S_{13}$ & $\bar{L}E$
& $(1,2,-{1\over 2})$
& ${1\over 2} g_1^2$
& 0.21
\\
$S_{14}$
& \multicolumn{2}{l}{
$N$ or $\bar{N}$ is included}
 & 0 & 0
\\
\hline
\end{tabular}
\end{center}
\end{table}
Each combination can contain other degrees of
freedom with respect to quantum numbers.
A large quantum number gives rise to
a large negative contribution which is 
the last term in Eq.~(\ref{c2}).
These degrees of freedom
are omitted in Table~\ref{tab:bs}.
For example, $\bar{Q}Q$ includes (1,3,0).
The binding strength for (1,3,0) is 
weaker than that of (1,1,0).

The field
$H_1$ has the quantum number for a Higgs doublet 
and $H_2$ corresponds to its dagger.
The fields
$S_1, \cdots, S_9$
have a mixed form for chirality 
such as the decomposition
$\bar{Q}_L Q_R$ and
$\bar{Q}_R Q_L$.
One of the decomposition, $Q_R$ for
$Q$ does not have zero mode.
Even if $\bar{Q}Q$ becomes a scalar bound state,
its potential would not give rise to
a negative mass squared.
The other combinations
$S_{10},\cdots S_{14}$ have zero modes
but binding strengths are relatively small.
If the net couplings of 
the constituents for $S_{12}$ and $S_{13}$ 
remain small,
the candidates of
composite scalars to potentially
yield a vacuum expectation value
would be only $H_1$ and $H_2$. 

For brane fermions, there are no correspondents
for $S_1,\cdots, S_9$.
Binding strengths for brane fermions
can be found in an analogous way to the list
given in Table~\ref{tab:bs}.

We assume that the ordinary gluon which is 
massless in four dimensions does not
lead to any condensation.
Therefore, our point for a condensation
is whether couplings are
clearly large compared to gluon couplings.
Numerical analysis for 
the four-dimensional effective
couplings will be performed 
in the next section.

\section{Numerical analysis for couplings \label{sec:coupling}}

The gauge couplings are included in 
covariant derivatives
in the action (\ref{action}).
The four-dimensional effective gauge interactions 
for zero-mode and KK-mode fields
are given by
\bea
 && \int d^4x 
  \sum_{n,m,\ell}
   \left(
     g_{L, nm\ell}^{\textrm{\scriptsize bulk}}
     \bar{\tilde{\Psi}}_{Ln}
       \gamma\cdot A_m 
       \tilde{\Psi}_{L\ell}
    +g_{R, nm\ell}^{\textrm{\scriptsize bulk}}
    \bar{\tilde{\Psi}}_{Rn} \gamma\cdot A_m
    \tilde{\Psi}_{R\ell}\right)
\nonumber
\\
  &\!\!\!+\!\!\!&
    \int d^4x \sum_{n,m,\ell}
    \left( g_{L,nm\ell}^{\textrm{\scriptsize brane}}
    \bar{\tilde{\Psi}}_{Ln} \gamma\cdot A_m
    \tilde{\Psi}_{L\ell}
    +g_{R,nm\ell}^{\textrm{\scriptsize brane}}
      \bar{\tilde{\Psi}}_{Rn} \gamma\cdot A_m
       \tilde{\Psi}_{R\ell} \right)
\nonumber
\\
  &\!\!\!+\!\!\!&
   \int d^4x \, \sum_{m}
   g_m \left(
    \bar{\hat{\Psi}}_L \gamma\cdot A_m P_c
    \hat{\Psi}_L
    +\bar{\hat{\Psi}}_R \gamma\cdot A_m P_c
    \hat{\Psi}_R \right) .
\eea
Here the four-dimensional effective gauge couplings
are given in terms of $z$-integral by
\bea
   g_{L,nm\ell}^{\textrm{\scriptsize bulk}}
   &\!\!\!=\!\!\!&
     {g_A\over \sqrt{L}} N_{Ln} N_{m} N_{L\ell}
      \int_1^{z_L}
        dz \, f_{Ln} \chi_m f_{L\ell} ,
\\
  g_{R,nm\ell}^{\textrm{\scriptsize bulk}}
   &\!\!\!=\!\!\!&
     {g_A\over \sqrt{L}} N_{Rn} N_{m} N_{R\ell}
      \int_1^{z_L}
        dz \, f_{Rn} \chi_m f_{R\ell} ,
\\
  g_{L,nm\ell}^{\textrm{\scriptsize brane}}
   &\!\!\!=\!\!\!&
    u_i k {g_A\over \sqrt{L}} N_{Ln} N_{m} N_{L\ell}
      \int_1^{z_L}
        dz \, f_{Ln} \chi_m f_{L\ell}
   \, z \delta(z-z_i) ,
     \label{branec1}
\\
  g_{R,nm\ell}^{\textrm{\scriptsize brane}}
   &\!\!\!=\!\!\!&
    u_i k {g_A\over \sqrt{L}} N_{Rn} N_{m} N_{R\ell}
      \int_1^{z_L}
        dz \, f_{Rn} \chi_m f_{R\ell}
   \, z \delta(z-z_i) ,
\\
  g_m &\!\!\!=\!\!\!&
    {g_A\over \sqrt{L}} N_m \int_1^{z_L}
      {dz \over z^3} \, \chi_m \delta(z-z_i) .
       \label{branec3}
\eea
We focus on
zero mode and the first three KK modes.
For each KK level, there are
the corresponding couplings
$g_{n0\ell}$, $g_{0m0}$, $g_{nmn}$, $g_{0mn}$,
$g_{nm\ell}$ where $n,m,\ell=1,2,3$
and the repetition of the same letter 
such as $g_{nmn}$ does not mean
the summation.
Bulk and brane couplings are given
for L- and R-even fermions in the following.

\subsubsection*{Bulk couplings}

For a zero-mode gauge boson,
the bulk couplings are given by
$g_{L,n 0\ell}^{\textrm{\scriptsize bulk}}
= g_{R, n 0 \ell}^{\textrm{\scriptsize bulk}}
= \delta_{n\ell}  g_A /\sqrt{L}$
where $n,\ell = 0,1,2,\cdots$.
The coupling of zero-mode fermions with a gluon 
is $g_A/\sqrt{L}$.
The condensation of our interest should have 
strong couplings compared to this value.
The summation of KK mode also
needs to be taken into account.

\begin{figure}[htb]
\begin{center}
\includegraphics[width=7.5cm]{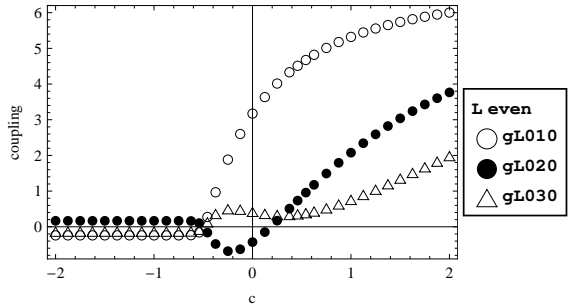}
\caption{Couplings 
$g_{L,0m0}^{\textrm{\scriptsize bulk}}$ 
divided by $g_A /\sqrt{L}$, where
the fermions are L even and 
the KK modes of the gauge bosons are $m=1,2,3$.
 \label{fig:gl0m0}}
\end{center}
\end{figure}

For zero-mode fermions, its coupling with
the first few KK-mode gauge bosons
is given in Fig.~\ref{fig:gl0m0}.\footnote{%
The overall sign 
for $g_{L,020}^{\textrm{\scriptsize bulk}}$
seems opposite to that of 
Ref.~\cite{Gherghetta:2000qt}.
This is due to the difference of the normalization.}
The figure is shown for L-even fermions.

For $c< -1/2$, the couplings 
are small and their $c$-dependence is small.
For $c> -1/ 2$, the size of the couplings
is rapidly enhanced.
The couplings with the first KK-mode 
tend to be large.
In a wide region for $c$, 
$|g_{L,010}^{\textrm{\scriptsize bulk}}|
>|g_{L,020}^{\textrm{\scriptsize bulk}}|
>|g_{L,030}^{\textrm{\scriptsize bulk}}|$.
The existence of an evident inequality
suggests the convergence of the summation
for KK modes.
In any case, for $c>-1/ 2$
the couplings become strong compared to
the gluon coupling.
For R-even fermions,
the couplings 
$g_{R,0m0}^{\textrm{\scriptsize bulk}}$
are given by
$g_{L,0m0}^{\textrm{\scriptsize bulk}}$
with the replacement 
$c\leftrightarrow -c$.

\begin{figure}[htb]
\begin{center}
\includegraphics[width=7.5cm]{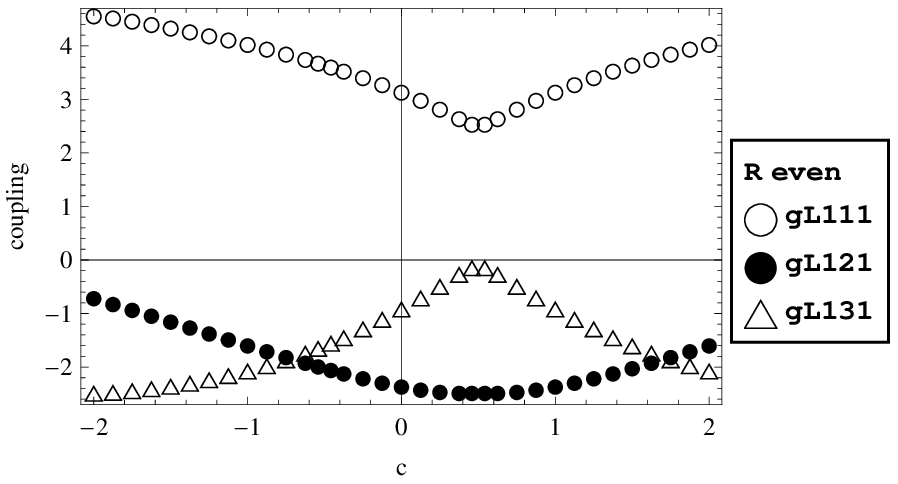} \qquad
\includegraphics[width=7.5cm]{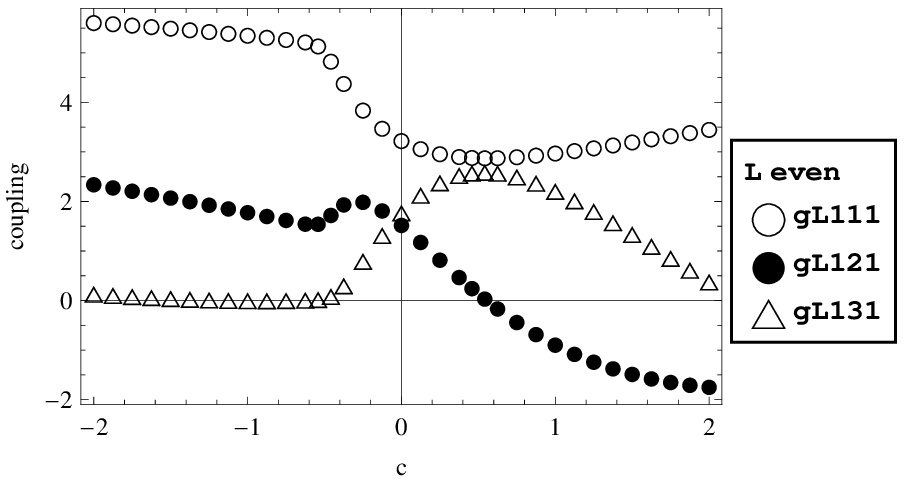}

\includegraphics[width=7.5cm]{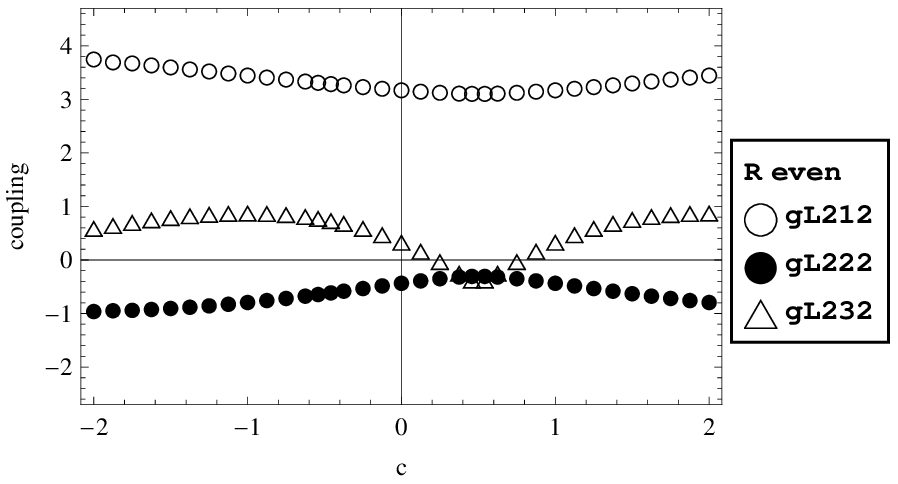} \qquad
\includegraphics[width=7.5cm]{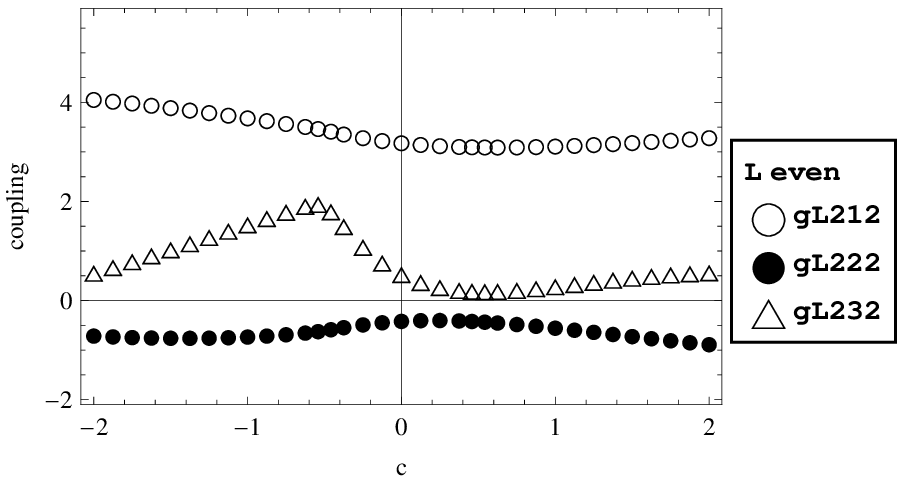}

\includegraphics[width=7.5cm]{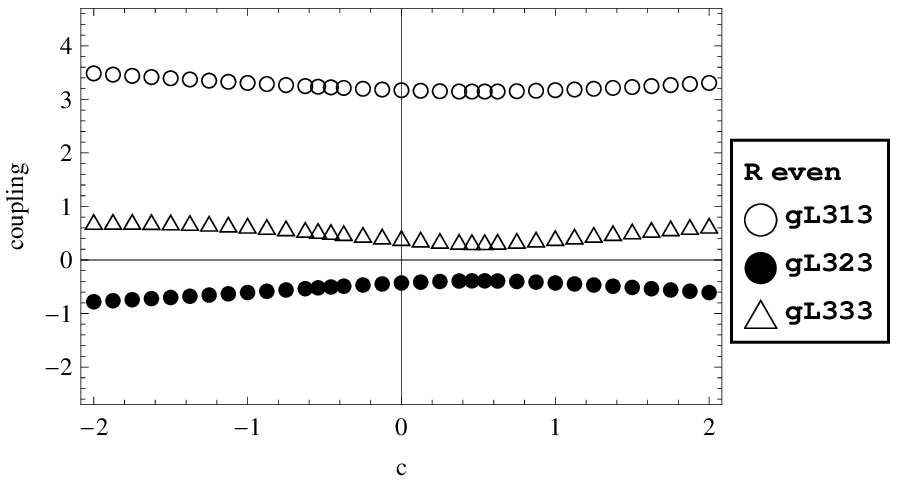} \qquad
\includegraphics[width=7.5cm]{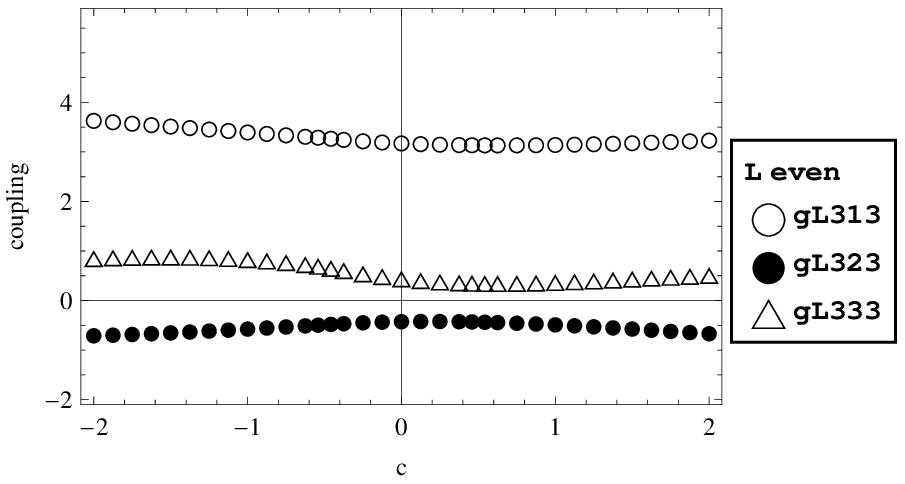}

\caption{Couplings 
$g_{L,nmn}^{\textrm{\scriptsize bulk}}$ 
divided by $g_A /\sqrt{L}$, where
the fermions
have identical KK modes $n=1,2,3$.
 \label{fig:gl1m1}}
\end{center}
\end{figure}

The bulk couplings with KK fermions, 
$g_{L,nmn}^{\textrm{\scriptsize bulk}}$ are
shown in 
Fig.~\ref{fig:gl1m1}.
Here the two fermions have identical KK levels.
For both of R-even and L-even fermions,
the couplings with the first KK gauge boson
are large independently of $c$, 
$g_{L,n1n}^{\textrm{\scriptsize bulk}} \simg 3$.
Unlike 
$g_{L,0n0}^{\textrm{\scriptsize bulk}}$
which significantly depends on $c$,
it seems inevitable that the couplings of
KK fermions with the first KK gauge boson 
are strong compared to
the gluon coupling.
On the other hand, it is read that
$|g_{Ln1n}^{\textrm{\scriptsize bulk}}|
> |g_{Ln2n}^{\textrm{\scriptsize bulk}}|,
|g_{Ln3n}^{\textrm{\scriptsize bulk}}|$.
This is an evident inequality similarly to the case of
$g_{L0n0}^{\textrm{\scriptsize bulk}}$.

From the couplings for left-handed fermions,
$g_{L,nmn}^{\textrm{\scriptsize bulk}}$,
the couplings for right-handed fermions,
$g_{R,nmn}^{\textrm{\scriptsize bulk}}$
can be obtained.
The couplings $g_{R,1m1}^{\textrm{\scriptsize bulk}}$,
$g_{R,2m2}^{\textrm{\scriptsize bulk}}$, 
$g_{R,3m3}^{\textrm{\scriptsize bulk}}$
for L-even fermions
have the same values as
$g_{L,1m1}^{\textrm{\scriptsize bulk}}$,
$g_{L,2m2}^{\textrm{\scriptsize bulk}}$,
$g_{L,3m3}^{\textrm{\scriptsize bulk}}$
for R-even fermions
with the replacement 
$c\leftrightarrow -c$, respectively where
$m=1,2,3$.
Similarly,
R-even $g_{R,1m1}^{\textrm{\scriptsize bulk}}$, $g_{R,2m2}^{\textrm{\scriptsize bulk}}$, 
$g_{R,3m3}^{\textrm{\scriptsize bulk}}$
have the same values as
L-even $g_{L,1m1}^{\textrm{\scriptsize bulk}}$,
$g_{L,2m2}^{\textrm{\scriptsize bulk}}$, 
$g_{L,3m3}^{\textrm{\scriptsize bulk}}$
with the replacement
$c\leftrightarrow -c$, respectively.

\begin{figure}[htb]
\begin{center}
\includegraphics[width=7.5cm]{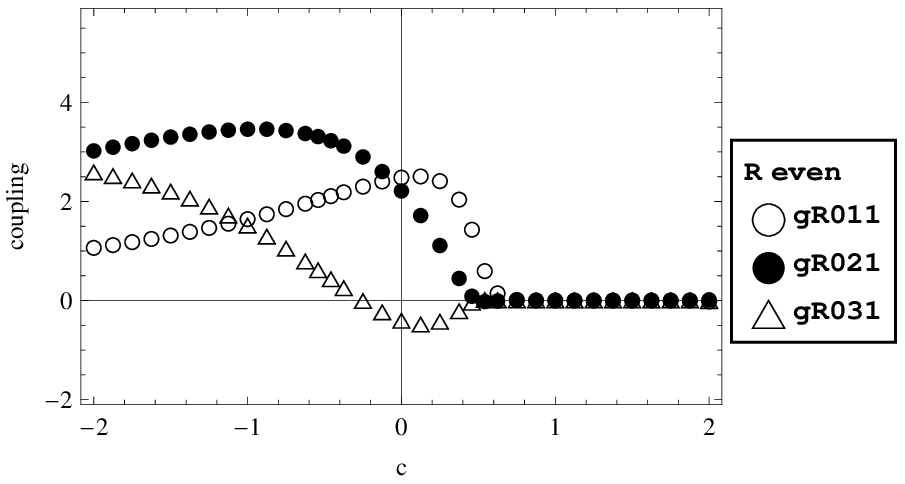} \qquad
\includegraphics[width=7.5cm]{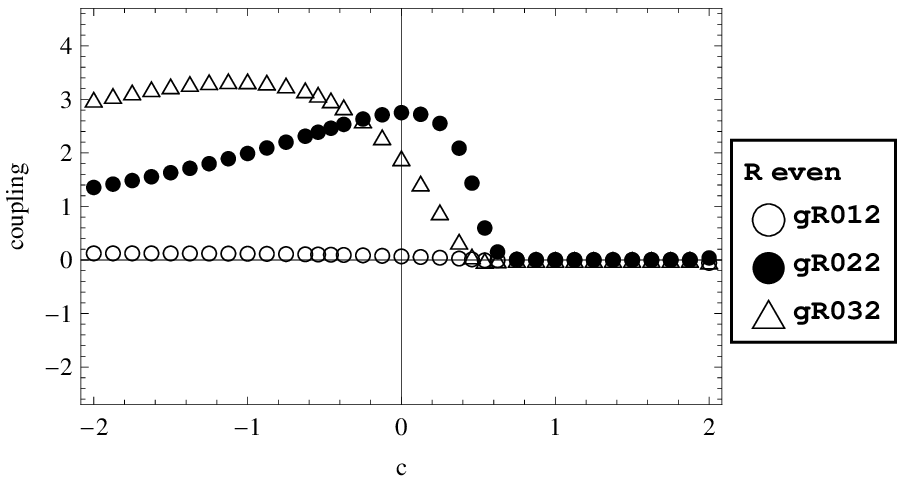}
\includegraphics[width=7.5cm]{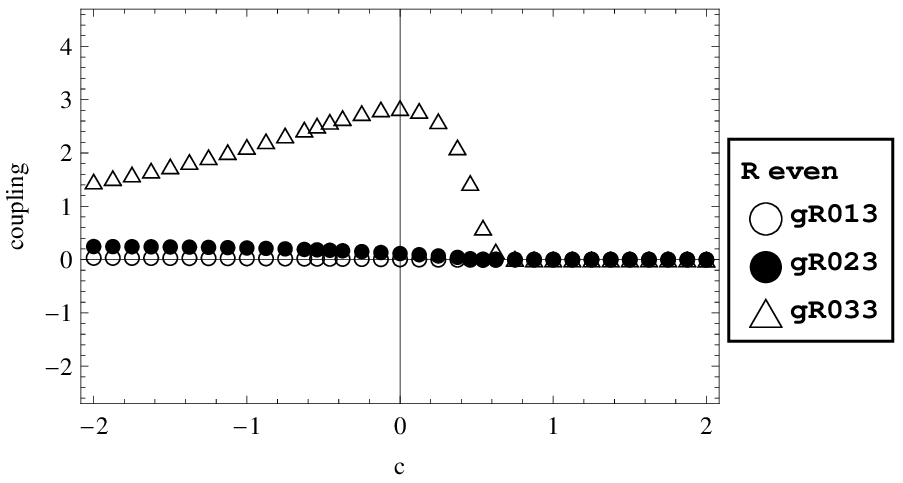}
\caption{Couplings 
$g_{R,0mn}^{\textrm{\scriptsize bulk}}$ 
divided by $g_A /\sqrt{L}$, where 
the fermions are R even. One fermion is zero mode
and the others are KK modes $n=1,2,3$.
 \label{fig:Re-gr0m1}}
\end{center}
\end{figure}

It is also possible that
only one of fermions in couplings is zero mode.
Zero modes belong to
the right-handed component for an R-even
fermion and the left-handed component for an L-even
fermion.
R-even 
$g_{L,0mn}^{\textrm{\scriptsize bulk}}$
and L-even
$g_{R,0mn}^{\textrm{\scriptsize bulk}}$
are vanishing
or do not exist where $n\neq 0$.
Non-vanishing couplings are shown in
Fig.~\ref{fig:Re-gr0m1}.
The couplings have a region with
a significant $c$-dependence and a region
almost independent of $c$. 
In this behavior, the couplings with
one zero-mode fermion resemble
$|g_{L,0m0}^{\textrm{\scriptsize bulk}}|$
and $|g_{R,0m0}^{\textrm{\scriptsize bulk}}|$
as we have seen in Fig.~\ref{fig:gl0m0}.

As $g_{L,010}^{\textrm{\scriptsize bulk}}
\neq 0$,
the KK number conservation is violated.
Furthermore, 
$g_{L,010}^{\textrm{\scriptsize bulk}}> g_{L,000}^{\textrm{\scriptsize bulk}}
=g_A/\sqrt{L}$ means that 
conserved quantities are not even
supported more than violated quantities.
However, among other couplings,
$|g_{R,022}^{\textrm{\scriptsize bulk}}|
> |g_{R,012}^{\textrm{\scriptsize bulk}}|$
and 
$|g_{R,033}^{\textrm{\scriptsize bulk}}|
> |g_{R,013}^{\textrm{\scriptsize bulk}}|,
|g_{R,023}^{\textrm{\scriptsize bulk}}|$
seem to support that
the KK number conservation is favored.
If this shows that there 
exists a critical $m$ for 
the largest coupling 
for each of 
$|g_{R,0m1}^{\textrm{\scriptsize bulk}}|$,
$|g_{R,0m2}^{\textrm{\scriptsize bulk}}|$,
$|g_{R,0m3}^{\textrm{\scriptsize bulk}}|$,
it may be related to the convergence for
the summation for KK modes. 
For $g_{R,0m0}^{\textrm{\scriptsize bulk}}$
and $g_{R,0mn}^{\textrm{\scriptsize bulk}}$,
it is interesting that the couplings
$|g_{R,010}^{\textrm{\scriptsize bulk}}|$,
$|g_{R,021}^{\textrm{\scriptsize bulk}}|$
and $|g_{R,032}^{\textrm{\scriptsize bulk}}|$,
in other words, 
$|g_{R,0(n+1)n}^{\textrm{\scriptsize bulk}}|$,
are large in a significantly $c$-dependent region.

For L-even fermions,
the couplings 
$g_{L,0mn}^{\textrm{\scriptsize bulk}}$
are given by the values with the same size 
and opposite sign as the R-even 
$g_{R,0mn}^{\textrm{\scriptsize bulk}}$
with the replacement $c\leftrightarrow -c$
where $m,n=1,2,3$.

\begin{figure}[htb]
\begin{center}
\includegraphics[width=7.5cm]{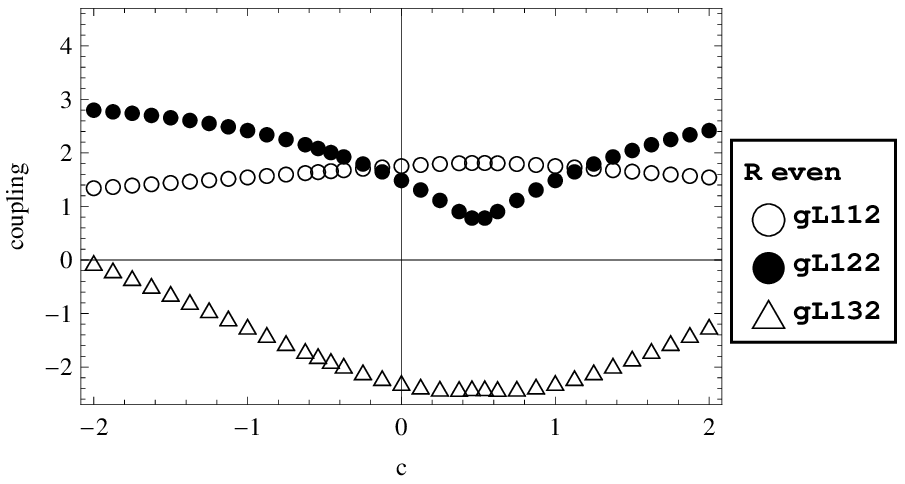} \qquad
\includegraphics[width=7.5cm]{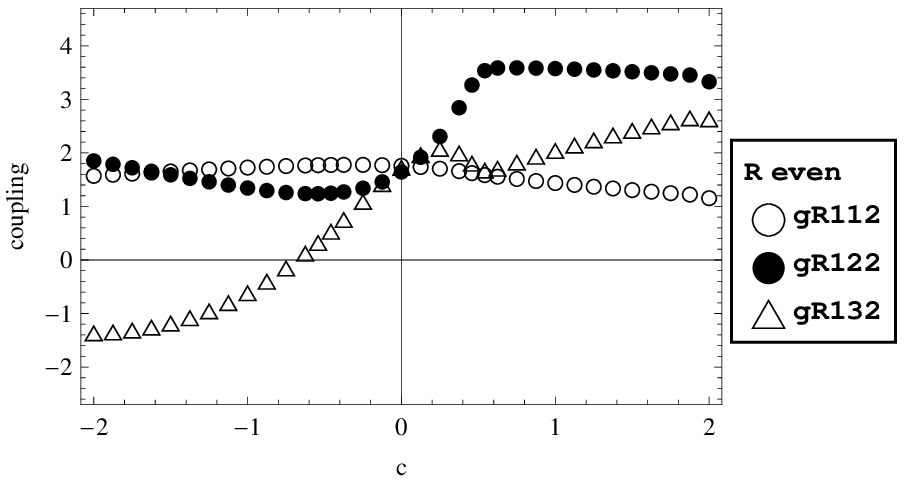}

\includegraphics[width=7.5cm]{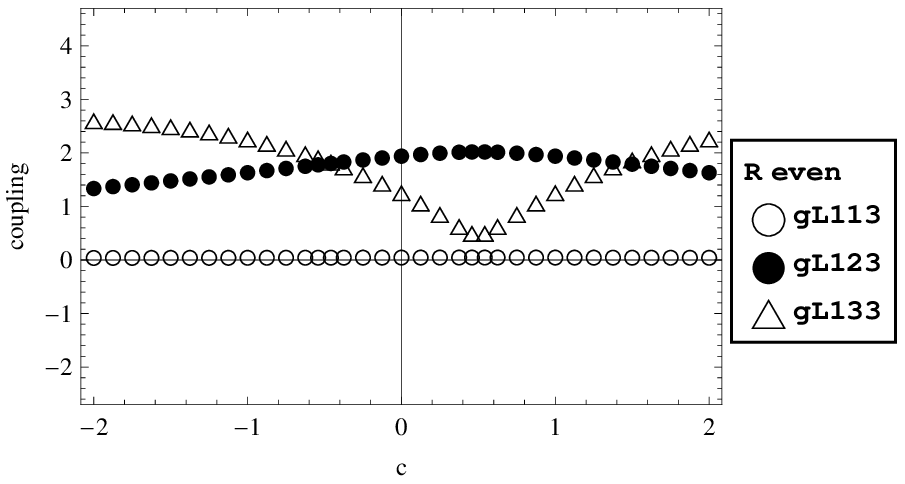} \qquad
\includegraphics[width=7.5cm]{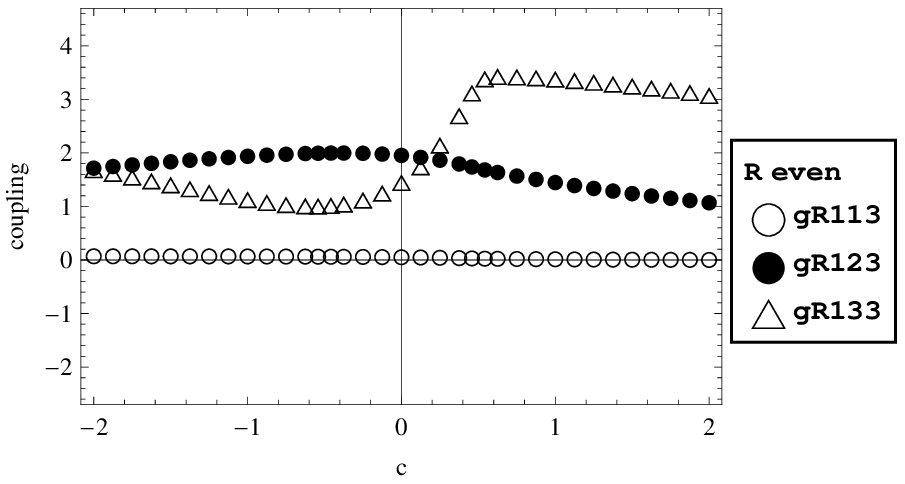}

\includegraphics[width=7.5cm]{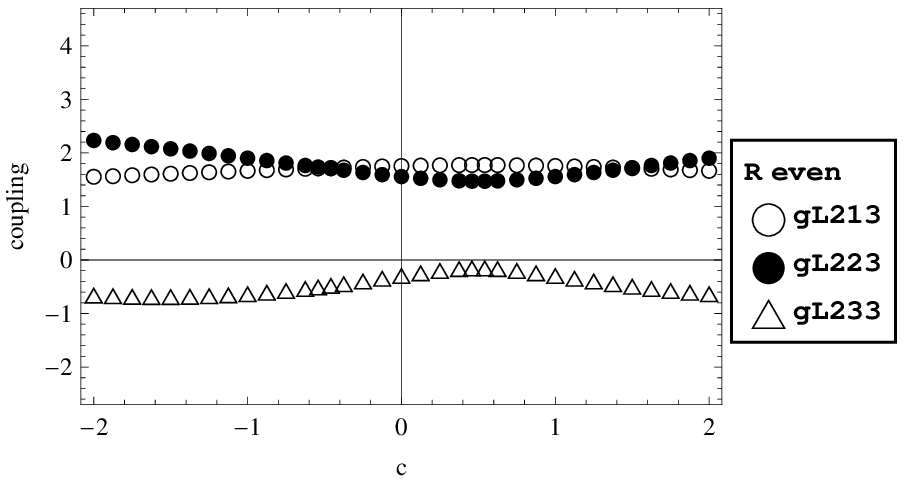} \qquad
\includegraphics[width=7.5cm]{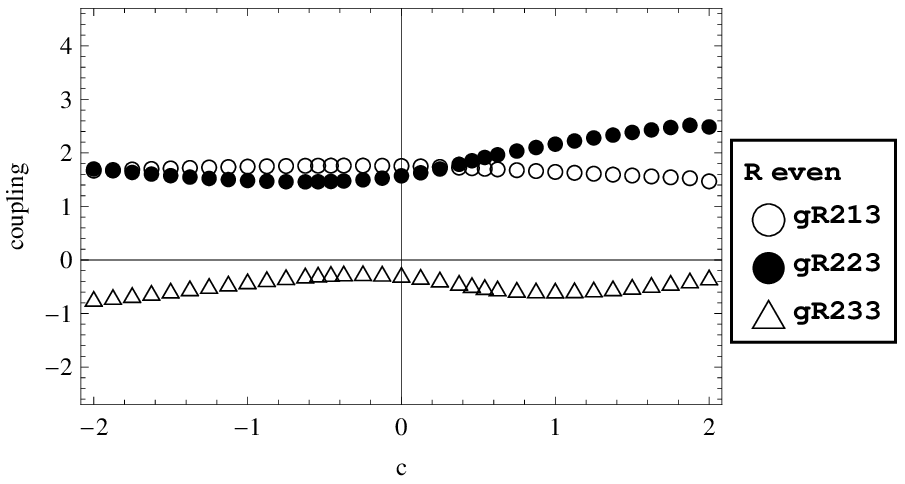}

\caption{Couplings 
$g_{L,nm\ell}^{\textrm{\scriptsize bulk}}$ 
and $g_{R,nm\ell}^{\textrm{\scriptsize bulk}}$ 
divided by $g_A /\sqrt{L}$, where
the fermions have different KK modes $n=1,2,3$
and $\ell = 1,2,3$.
 \label{fig:gl1m2}}
\end{center}
\end{figure}

The other couplings are KK gauge couplings for
fermions with different KK modes.
The $g_{L,nm\ell}^{\textrm{\scriptsize bulk}}$
and $g_{R,nm\ell}^{\textrm{\scriptsize bulk}}$
with $n,m,\ell =1,2,3$ and $n\neq \ell$ 
are shown for R-even fermions
in Fig.~\ref{fig:gl1m2}.  
The profile 
of R-even $g_{L,nm\ell}^{\textrm{\scriptsize bulk}}$
resembles
that of R-even $g_{L,nmn}^{\textrm{\scriptsize bulk}}$.
For R-even fermions, in wide regions 
$|g_{L,122}^{\textrm{\scriptsize bulk}}|
>|g_{L,132}^{\textrm{\scriptsize bulk}}|
>|g_{L,112}^{\textrm{\scriptsize bulk}}|$,
$|g_{L,133}^{\textrm{\scriptsize bulk}}|
>|g_{L,123}^{\textrm{\scriptsize bulk}}|
>|g_{L,113}^{\textrm{\scriptsize bulk}}|$,
$|g_{L,223}^{\textrm{\scriptsize bulk}}|
>|g_{L,213}^{\textrm{\scriptsize bulk}}|
>|g_{L,233}^{\textrm{\scriptsize bulk}}|$.
All the largest couplings above fulfill
\bea
  \ell -n +1 =m ,
   \label{levelrelation}
\eea
for the KK levels.
This relation is fulfilled 
also for 
$g_{R,0(n+1)n}^{\textrm{\scriptsize bulk}}$
which are large couplings as found above.
It is remarkable that 
the KK level relation given in
Eq.~(\ref{levelrelation})
is not the KK mode conservation.

The values of the couplings can be larger
than that of the gluon coupling 
independently of $c$.  
This behavior is the same for
$g_{R,nm\ell}^{\textrm{\scriptsize bulk}}$.
For L-even fermions,
$g_{L,nm\ell}^{\textrm{\scriptsize bulk}}$
and $g_{R,nm\ell}^{\textrm{\scriptsize bulk}}$
is given by
$g_{R,nm\ell}^{\textrm{\scriptsize bulk}}$
and $g_{L,nm\ell}^{\textrm{\scriptsize bulk}}$
with the replacement $c\leftrightarrow -c$,
respectively.  
 
\vspace{2ex}

In summary for bulk couplings, 
the ordinary gluon, or the four-dimensional
massless gluon  couples to fermions 
with at most $g_A/\sqrt{L}$ where
fermions are not only zero mode but also KK modes.
Therefore, the zero-mode gluon is not 
the dominant part for a condensation
even if the components of fermions are KK modes.
A KK-mode gluon can have large couplings 
compared to $g_A/\sqrt{L}$.
When at least one of fermions is zero mode,
the gauge couplings can be small.
For L-even fermions with $c< -1/2$,
the KK gauge couplings have
$g_{L,0mn}^{\textrm{\scriptsize bulk}}
< g_A/\sqrt{L}$ where $m=1,2,3$
and $n=0,1,2,3$.
For R-even fermions with $c> 1/2$,
the KK gauge couplings have
$g_{R,0mn}^{\textrm{\scriptsize bulk}}
 < g_A /\sqrt{L}$.
Hence,
for L-even fermions with $c<-1/2$
and R-even fermions with $c>1/2$,
the zero modes of a gluon and fermions
do not give rise to a condensation solely 
even if the KK modes of the gluon and fermions
are included in the couplings.
Without zero modes,
KK gauge couplings such as 
$g_{L,111}^{\textrm{\scriptsize bulk}}$
have large values compared to $g_A/\sqrt{L}$
independently of $c$.
In other words, a condensation may be
an inevitable effect composed of KK-mode fields.

This evaluation is applied to the binding strengths
given in Table~\ref{tab:bs}.
The maximum absolute value of the 
couplings is about $6 g_A/\sqrt{L}$ 
for $-2<c<2$.
Even after the contributions are summed with
respect to KK modes,
$S_{12}$ may be small on account of
the smallness of the relative strength.
We will see that the constraint for 
$S_{12}$ mainly arises from the brane coupling
rather than the bulk coupling.
On the other hand, $S_{13}$ can have large values.
The quantum number of $S_{13}$ is the same as 
that of $H_2$.
It may yield a mixing.
For bulk fields, SU(3)-singlet and SU(2)-doublet
$H_1$, $H_2$ and $S_{13}$
are candidates of composite scalars.

\subsubsection*{Brane couplings}

The boundary values of the mode functions
have been found
in the end of Sec.~\ref{sec:mf}.
The brane couplings (\ref{branec1})-(\ref{branec3})
are given by the products of 
the boundary values of the mode functions.
We write down the couplings for L-even fermions
in the following.
For R-even fermions, the couplings
are given by the corresponding 
couplings for L-even fermions with 
the replacement
$c\leftrightarrow -c$.

\begin{figure}[htb]
\begin{center}
\includegraphics[width=7.5cm]{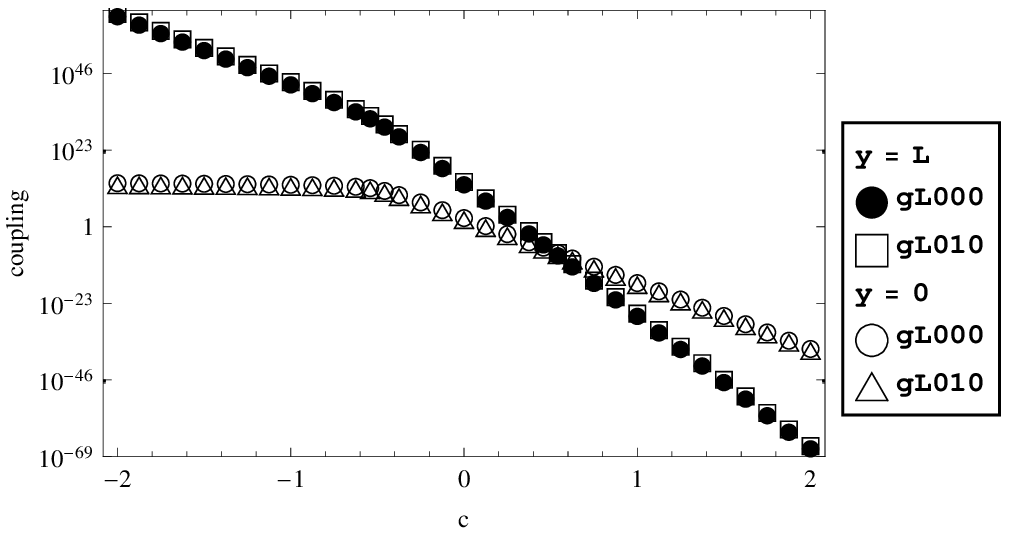} \qquad
\includegraphics[width=7.5cm]{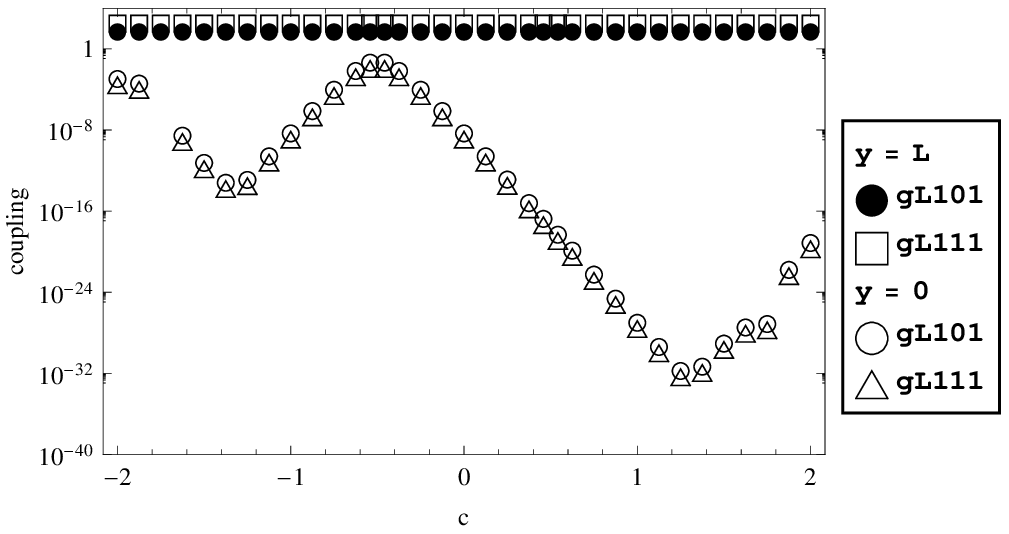}

\includegraphics[width=7.5cm]{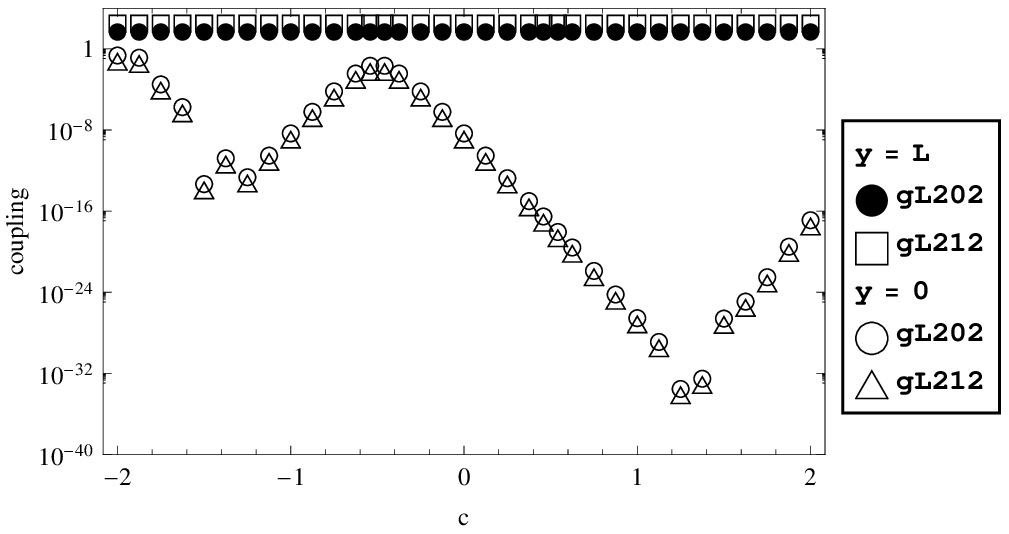} \qquad
\includegraphics[width=7.5cm]{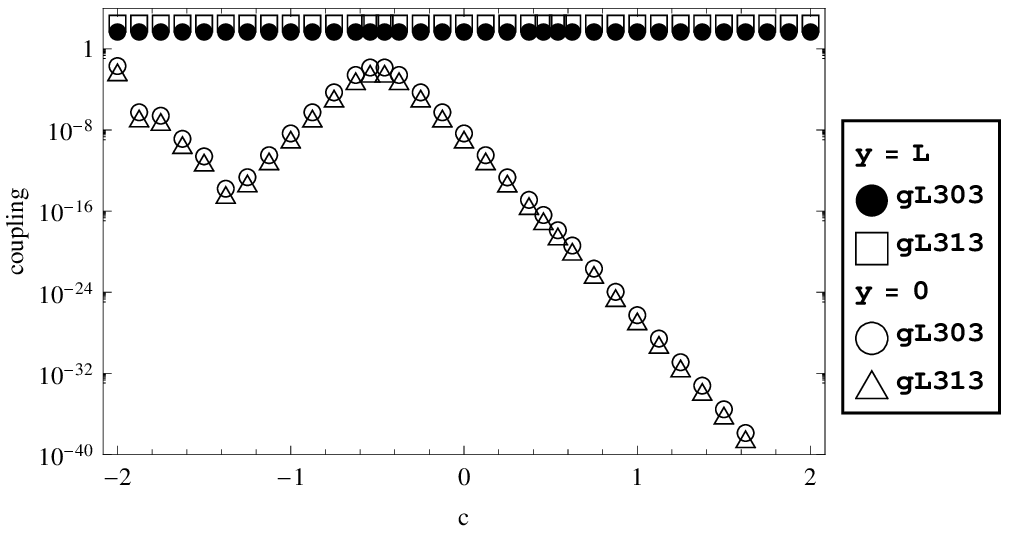}
\caption{Coupling 
$g_{L,nmn}^{\textrm{\scriptsize brane}}$
divided by $(u_i/L) g_A /\sqrt{L}$,
where the fermions are L even and
have identical 
KK modes $n=0,1,2,3$. \label{fig:b0m0}}
\end{center}
\end{figure}

For L-even fermions, the $c$-dependence of
$g_{L,nmn}^{\textrm{\scriptsize brane}}$
divided by $(u_i/L) g_A /\sqrt{L}$
is shown in Fig.~\ref{fig:b0m0}.
The values of $g_{L,n0n}^{\textrm{\scriptsize brane}}$
and $g_{L,n1n}^{\textrm{\scriptsize brane}}$
are not sensitive to the level of KK modes for fermions.
As seen in Table~\ref{tab:chi},
this is because 
the boundary values $N_n \chi_n\bigg|$ 
depend on the KK level weakly.
From the values in Table~\ref{tab:chi},
the values of 
$g_{L,n2n}^{\textrm{\scriptsize brane}}$
and $g_{L,n3n}^{\textrm{\scriptsize brane}}$
can also be obtained.

For $c\simg 1/2$, the couplings
$g_{L,0m0}^{\textrm{\scriptsize brane}}$
are small, 
$g_{L,0m0}^{\textrm{\scriptsize brane}}
< (u_i/L) g_A /\sqrt{L}$
 at both of $y=0$ and $y=L$ where 
$m=0,1,2,3$.
In particular, the brane coupling
of zero modes of L-even fermions
with zero mode of a gluon is
given by
$g_{L,000}^{\textrm{\scriptsize brane}}$.
The value at $y=L$ can be very large for fermions 
with $c\siml 1/2$, depending on $u_i$.
A scenario to
avoid a condensation by the ordinary gluon
may require that
the bulk mass parameter
for all L-even fermions
should be taken as $c\simg 1/2$.

When fermions are KK modes,
$g_{L,nmn}^{\textrm{\scriptsize brane}}$
are small for $y=0$
and large for $y=L$ where
$n=1,2,3$.
If $u_i$ do not include suppression factors,
the massless gluon strongly 
couples to KK modes of fermions at $y=L$.
This property is almost independent of $c$.

\begin{figure}[htb]
\begin{center}
\includegraphics[width=7.5cm]{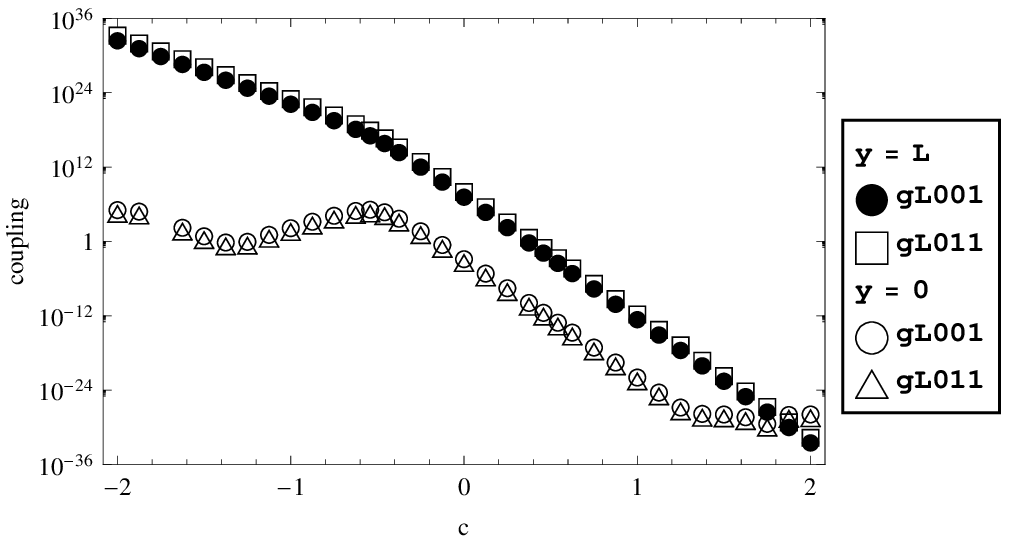} \qquad
\includegraphics[width=7.5cm]{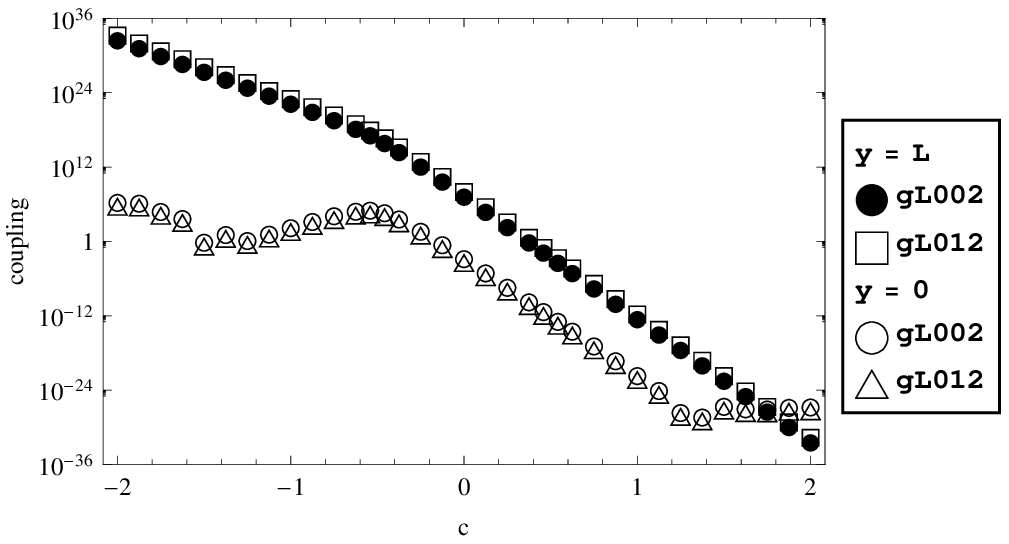}

\includegraphics[width=7.5cm]{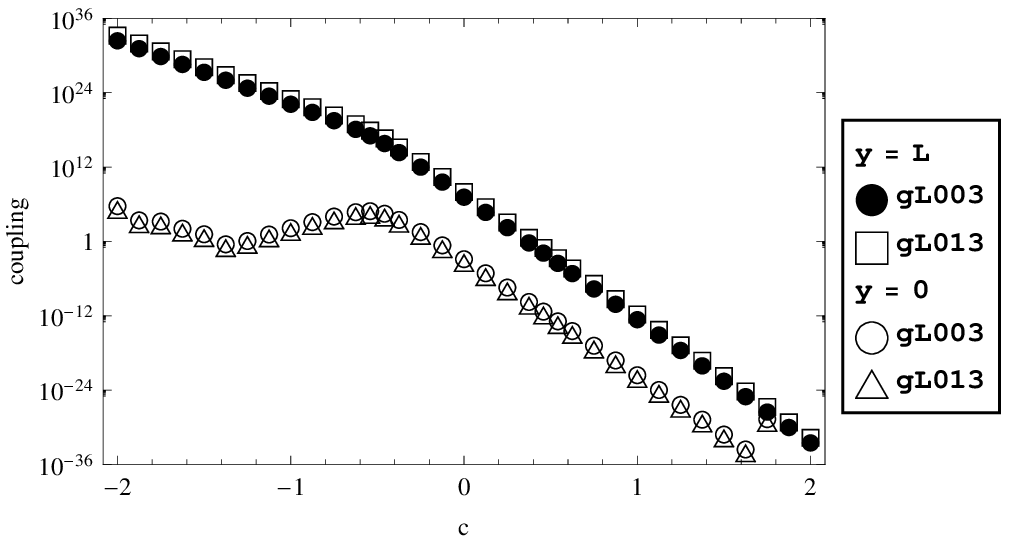} \qquad
\includegraphics[width=7.5cm]{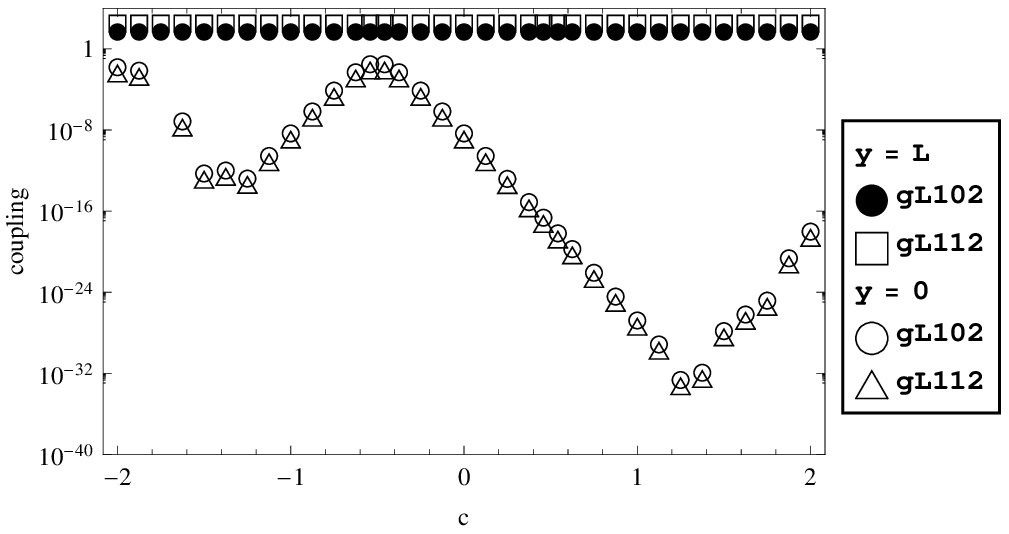}

\includegraphics[width=7.5cm]{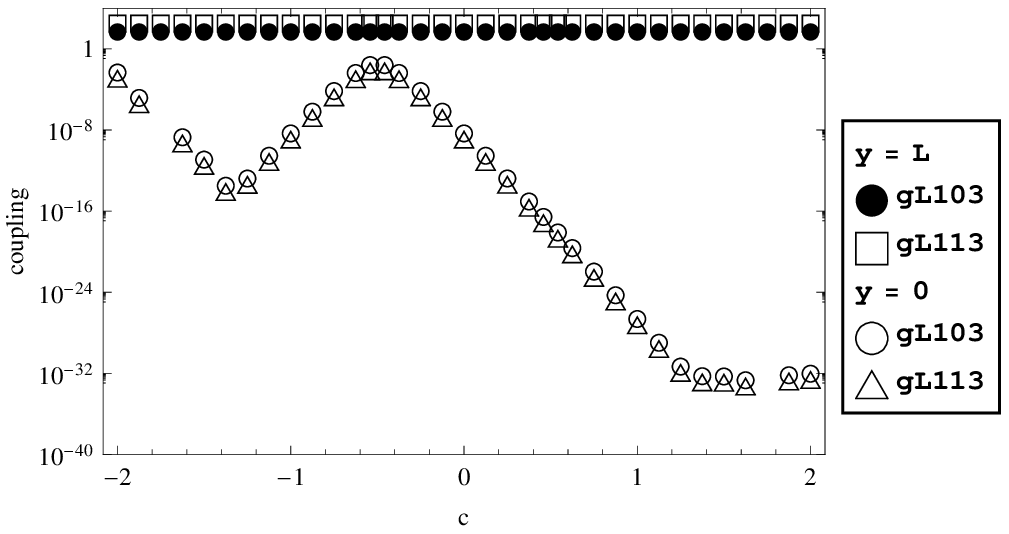} \qquad
\includegraphics[width=7.5cm]{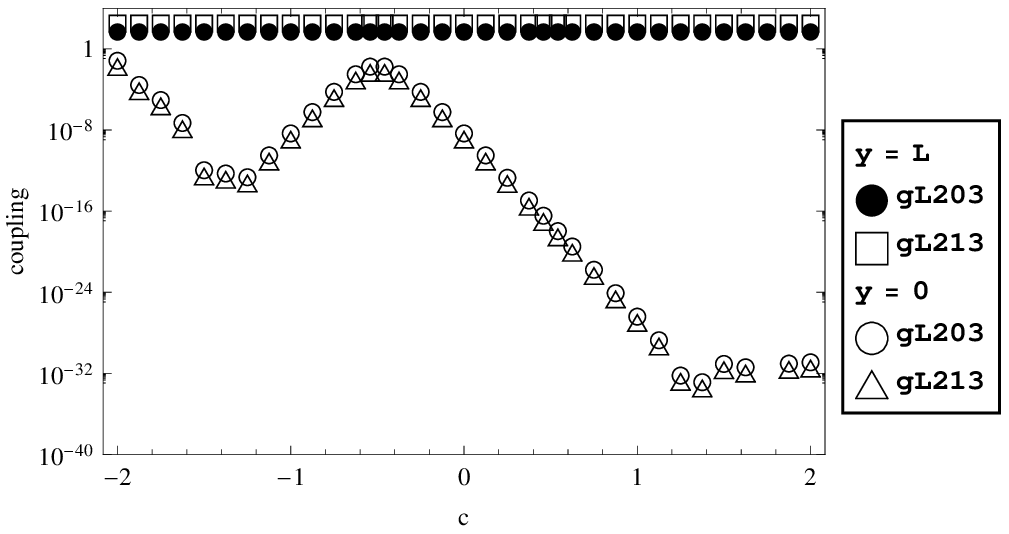}
\caption{Coupling 
$g_{L,nm\ell}^{\textrm{\scriptsize brane}}$
divided by $(u_i/L) g_A /\sqrt{L}$,
where the fermions are L even and
have different 
KK modes $n=0,1,2,3$ and $\ell = 0,1,2,3$. 
  \label{fig:b0m1}}
\end{center}
\end{figure}

Also in the case where 
KK-mode fermions are not identical levels,
gauge couplings are non-vanishing.
The couplings
$g_{L,nm\ell}^{\textrm{\scriptsize brane}}$
divided by $(u_i/L) g_A /\sqrt{L}$,
are shown in Fig.~\ref{fig:b0m1}.
where the fermions are L even and
have different 
KK modes $n=0,1,2,3$ and $\ell = 0,1,2,3$.

The $c$-dependence of the couplings
$g_{L,0mn}^{\textrm{\scriptsize brane}}$
are similar to that of
$g_{L,0m0}^{\textrm{\scriptsize brane}}$ in
Fig.~\ref{fig:b0m0}
where $m=0,1,2,3$ and $n=1,2,3$.
The $c$-dependence of the couplings
$g_{L,0mn}^{\textrm{\scriptsize brane}}$
are similar to that of
$g_{L,0m0}^{\textrm{\scriptsize brane}}$ in
Fig.~\ref{fig:b0m0}
where $m=0,1,2,3$ and $n=1,2,3$.

Unlike the bulk couplings,
the brane couplings have 
a small dependence of the KK level
for the gauge boson.
The summation of the KK mode 
can yield a large contribution.

Lastly the brane coupling $g_m$ in 
Eq.~(\ref{branec3}) are given by the products
of the boundary values
$N_m \chi_m\bigg|$ and $z_i^{-3}$.
At $y=0$, the values of $\chi_m\bigg|$
are positive and negative alternately
with respect to the KK level. 
In addition, the absolute value at each KK level
is small. 
The coupling $g_m$ for $y=0$ 
seems small compared to other 
typical bulk and brane couplings.
At $y=L$, 
the couplings are multiplied by $z_L^{-3}$.
The coupling $g_m$ also for $y=L$ 
seems small.
Therefore bound states formed
by brane fermions such as $\bar{\hat{Q}}\hat{U}$ 
tend to have small attractive forces
compared to that of bulk fermions.

It has been found that 
bulk and brane couplings notably differ
in the dependences on
$c$ and the KK level.
The small bulk couplings for L-even fermions are
summarized as
$g_{L,n0\ell}^{\textrm{\scriptsize bulk}}$ 
for any $c$ and
$g_{L,0mn}^{\textrm{\scriptsize bulk}}$ for 
$c\siml -1/2$
where $n,\ell=0,1,2,3$ and $m=1,2,3$.
The small brane couplings for L-even fermions are
summarized as
$g_{L,0mn}^{\textrm{\scriptsize brane}}$
for $c\simg 1/2$
where $n,m=0,1,2,3$
and $g_{L,nm\ell}^{\textrm{\scriptsize brane}}$
at $y=0$ for $c\simg 1/2$
where $n,\ell=1,2,3$ and $m=0,1,2,3$.
Here the convergence of brane couplings 
with respect to the summation
of KK modes seems worse than that of bulk couplings. 
For R-even fermions,
the discussion is parallel
with $c\leftrightarrow -c$.

\subsubsection*{A scenario for a condensation
to trigger electroweak symmetry breaking}
 
We assume that
zero-mode fermions and a zero-mode gauge boson
are not condensed.
The bulk coupling 
 is $g_{000}^{\textrm{\scriptsize bulk}}
=g_A/\sqrt{L}$ for any $c$.
The brane couplings
$g_{L,000}^{\textrm{\scriptsize brane}}$
and $g_{R,000}^{\textrm{\scriptsize brane}}$
depend on $c$.
The mass parameter to satisfy this condition
is $c \simg 1/2$ for L-even fermions
and $c \siml -1/2$ for R-even fermions.
 
For $c\simg 1/2$, the coupling with
a KK-mode gauge boson is large,
$g_{L,010}^{\textrm{\scriptsize bulk}} 
\simg 4 g_A/\sqrt{L}$.
Even zero-mode fermions
receive strong attractive force through
a KK-mode gauge boson.
Also,
even a zero-mode gauge boson
can have large couplings with KK-mode fermions at $y=L$.
When
some of fermions and gauge bosons
are zero modes and the others are KK modes, 
the couplings become strong.

A substantial
point to realize electroweak symmetry breaking
is that 
SU(3)-triplet
$S_{12}$ in Table~\ref{tab:bs}
has a zero vacuum expectation value.
A simple way to achieve this
is to avoid a strong coupling for
$S_{12}$.
As the relative strength is 0.07,
the bulk couplings whose maximum is about 
$6g_A/\sqrt{L}$ seem to remain small.
The small brane couplings need at least
$c\simg 1/2$ for L-even fermions. 
From the viewpoint of both of
the zero-mode coupling and the $S_{12}$
coupling, the mass parameters
$c\simg 1/2$ for L-even fermions
and $c\siml -1/2$ for R-even fermions
are favored.
This tendency is the same also for 
brane gauge couplings with 
one bulk fermion and one brane fermion
due to the $\Psi$-$\hat{\Psi}$ mixing.
Correspondingly, the contributions such as
$\bar{\Psi}_{1L0} \Psi_{2R0}$ 
are localized at the brane with an 
intermediate-scale cutoff as described 
in Sec.~\ref{sec:mf}.

When $S_{12}$ does not 
have a vacuum expectation value,
SU(2)-doublet scalars are only candidates
for non-vanishing vacuum expectation values.
The SU(2)-doublet scalars can be condensed
through constituents with
large attractive contributions from KK modes, 
while
dynamical degrees of freedom for
zero-mode fermions and zero-mode gluon
are kept.

\section{Boson masses and weak mixing angle
\label{sec:bose}}

When electroweak symmetry breaking occurs,
gauge boson and Higgs boson acquire their masses.
We estimate how
the gauge boson mass, weak mixing angle and Higgs 
boson mass are described in terms of 
the possible vacuum expectation values.

For simplicity, we 
focus on 
the composite Higgs $H_1(x,z)$ whose constituents
are $\bar{Q}U(x,z)$.
The extension for inclusion of 
$H_2$ and $S_{13}$ is straightforward.
Now the notation for a bulk composite Higgs
is denoted as $H$. 
The inclusion of brane fields such 
as $\hat{H}_1$ will be shown explicitly.

The composite Higgs $H(x,z)$ has
the interactions
\bea
  && \int d^4 x dz \, \sqrt{\textrm{det}(g_{KL})}
    \, g^{MN} (D_M H)^\dag (D_N H) 
\nonumber
\\
  &\!\!\!+\!\!\!&
    a_i \int d^4 x dz \sqrt{\textrm{det}(g_{KL})}
       \, g^{\mu\nu} (D_\mu H)^\dag (D_\nu H)
   kz \delta (z-z_i) 
\eea
where the overall factor has been normalized
for the bulk kinetic term.
For the present analysis, we regard $a_i$ as 
parameters.
The covariant derivatives includes 
the gauge couplings as
$D_M H \supset 
(-ig_2 W_M -i (1/ 2) g_1 B_M ) H$
where $Y_{\bar{Q}U}= 1/2$.
The zero modes of gauge fields are 
given by
$(1/\sqrt{L}) W_\mu^{(0)}$
and $(1/\sqrt{L}) B_\mu^{(0)}$.

When the condensation occurs,
some energy would be released.
Originally the fields $Q_L$ and $U_R$ 
have massless modes.
The scalar bound state can fall on the minima of 
its potential.
When the vacuum expectation value is generated as
$\langle H \rangle = 
 (0, v^{3/2} )^T$, 
the light gauge fields have the mass terms as
\bea
   \int d^4x \, 
  {v_{\textrm{\scriptsize eff}}^2 \over 2}
     \left[
       {g_2^2 \over L}
         \left(
          {(W_\mu^{(0)1})^2 + (W_\mu^{(0)2})^2
            \over 2}\right)
     +{1\over 2} {g_1^2 + g_2^2 \over L}
            \left({- g_2 W_\mu^{(0)3}
         + g_1 B_\mu^{(0)}
 \over \sqrt{g_1^2+g_2^2}} \right)^2 \right] ,
\eea
with
$v_{\textrm{\scriptsize eff}}^2
   =    \left[ 
        (1- z_L^{-2})/(2k)
 + a_0 + a_1  z_L^{-2} \right] v^3$.
The weak mixing angle is
given similarly to the standard model,
$\cos \theta_W = m_W/m_Z
=g_2/\sqrt{g_1^2 +g_2^2}$.
It is independent of $v_{\textrm{\scriptsize eff}}$.

If the brane fields add the effect of a condensation
corresponding to
$\langle \hat{H} \rangle =
   ( 0 , 
      \hat{v})^T$,
the effective expectation value
 $v_{\textrm{\scriptsize eff}}^2$ 
is replaced with
$v_{\textrm{\scriptsize eff}+}^2
   = v_{\textrm{\scriptsize eff}}^2  
+ \hat{v}^2$.
The weak mixing angle is the same as 
the case of $\hat{v}=0$.

The couplings $g_2/\sqrt{L}$ and $g_1/\sqrt{L}$ are 
effective four-dimensional couplings.
In order that the gauge bosons have 
the masses at the weak scale,
the $v_{\textrm{\scriptsize eff}}$ should be of the order of 
${\cal O}(100)$GeV. 
For $k=4\times 10^{12}$GeV
and $a_0\sim a_1 \sim L$,
the definition of $v_{\textrm{\scriptsize eff}}$ means 
$v^{3/2}\sim {\cal O}(10^7)$GeV.

From a four-dimensional 
standard representation for Higgs mass
$\lambda v^2 = m_H^2$,
the present Higgs mass is estimated as
$\lambda v^3 = m_H^2$ where $\lambda$ is
the coefficient for a fermion four-point interaction.
When strongly-coupled effects
for gauge interactions are denoted as $N_{n p}$,
the coupling is given by
$\lambda \sim
    g^2 N_{n p} 
  \sim 
    g_4^2 L N_{n p} 
  \sim
    g_4^2 (\log z_L /k) N_{n p}$.
Then the Higgs mass squared is given by
$m_H^2 = g_4^2 \log z_L \cdot N_{n p} 
   \, v^3/k
\sim g_4^2 N_{n p} \cdot {\cal O}(1000)$GeV.
In the present setup,
the KK fields are heavy. 
As shown in Table~\ref{tab:kkmass},
the lightest mass is about 10 TeV.
In order that the tree level unitarity is not 
violated,
the Higgs boson may need to contribute
below the scale where KK fields become dynamical.
If $N_{n p}$ is not extremely large,
the renormalization
group running 
from the scale $v^{3/2}$ to the weak scale
may drive
the Higgs boson mass to ${\cal O}(100)$GeV.

\section{Conclusion \label{conclusion}}

We have systematically 
examined gauge couplings for fermions in 
a warped space. 
Here bulk and brane coupling of bulk fermions
and brane coupling of brane fermions
have been calculated.
This has been related to binding strengths
in the most-attractive-channel approximation.

From the quantum number given, 
there are six color-triplet scalars for
one-generation bulk fermions.
For the three scalars $S_7$, $S_8$, $S_9$, 
one of the constituents
$\bar{\Psi}_1$ and $\Psi_2$
is necessarily a KK mode.
The two scalars,
$S_{10}$ and $S_{11}$ do not receive attractive force.
Then the dangerous color-triplet scalar is 
only $S_{12}$.
Brane fermions yields the correspondent $\hat{S}_{12}$.
The scenario is 
that these bound states are
prevented from having their
vacuum expectation values
while weak-doublet scalar bound states 
have non-vanishing vacuum expectation values 
to trigger symmetry breaking.

The boundary values of the mode functions
significantly depend on the bulk mass parameters
for zero-mode fermions.
The $c$-dependence of 
KK-mode fermions and both modes of gauge bosons
is small.
Therefore, the brane couplings
for zero-mode fermions most strongly depend
on $c$.
To avoid large couplings of zero-mode fermions
such as a zero-mode gluon-fermion-fermion
yields constraints on $c$. 
The bulk and brane gauge couplings of bulk fermions
differ in the convergence of the KK summation.
From this point of view,
the brane gauge couplings are 
worse than the bulk couplings.
The contributions of the brane gauge coupling
at each KK level can be small
depending on the bulk mass parameter.
The brane gauge couplings of brane fermions
tend to be smaller than the brane gauge couplings
of bulk fermions.
The bulk coupling for zero-mode gluon and fermions
is given by
$g_{000}^{\textrm{\scriptsize bulk}} =g_A/\sqrt{L}$
for any $c$.
The condition that brane couplings 
$g_{L,000}^{\textrm{\scriptsize brane}}$
and $g_{R,000}^{\textrm{\scriptsize brane}}$
are at most $\sim g_A/\sqrt{L}$ 
gives
$c\siml 1/2$ for L-even fermions and 
$c\simg-1/2$ for R-even fermions.
The condition that the
couplings for the color-triplets are
at most $\sim g_A/\sqrt{L}$ 
requires the same region for $c$.

We have found an interesting property
for the bulk couplings.
For $\bar{\tilde{\Psi}}_n \gamma \cdot A_{m} 
\tilde{\Psi}_\ell$,
large couplings are
given for
$\ell -n+1=m$ in Eq.~(\ref{levelrelation}). 
It is curious that this relation 
is not to represent the conservation of 
KK modes.
The well-known
$g_{L,010}^{\textrm{\scriptsize bulk}}
>g_{000}^{\textrm{\scriptsize bulk}}$
in a wide region of $c$
is characteristic.
In particular, for $c\simg 1/2$
the KK gluon coupling
is clearly large
$g_{L,010}^{\textrm{\scriptsize bulk}}
\simg 5 g_{000}^{\textrm{\scriptsize bulk}}$.
Large couplings of KK gluon are a necessary 
consequence
when the strong coupling for a zero-mode gluon 
is avoided.

The gauge boson masses, weak mixing angle 
and Higgs boson masses have
been related to the vacuum expectation values.
The weak mixing angle is written in terms
of the coupling constants as in the standard model.
For $k=4\times 10^{12}$GeV and 
$a_0 \sim a_1 \sim L$,
the vacuum expectation value has been
estimated as
$v^{3/2} \sim {\cal O}(10^7)$GeV.
As the KK fields are heavy and the lightest mass
is about 10TeV,
the Higgs boson may need to be lighter than them
so that
unitarity is not violated.
The renormalization group flow
from the high scale $v^{3/2}$ to the weak scale
may make the Higgs boson mass light.

We have examined the aspect of the gauge couplings
for the self-breaking of
the standard model gauge symmetry
in the warped space.
Many issues such as
the actual vacuum expectation values
and the flavor mixing
need to be examined in more detail. 
This and the comparison
with experimental data would require
some fundamental development of estimation for
quantum corrections
with extra dimensions.

\vspace{8ex}

\subsubsection*{Acknowledgments}

This work is supported by Scientific Grants 
from the Ministry of Education
and Science, Grant No.~20244028.





\vspace*{10mm}



\begin{thebibliography}{99}



\bibitem{ArkaniHamed:2000hv}
  N.~Arkani-Hamed, H.~C.~Cheng, B.~A.~Dobrescu and L.~J.~Hall,
  Phys.\ Rev.\  D {\bf 62}, 096006 (2000)
  [arXiv:hep-ph/0006238].



\bibitem{Dobrescu:1998dg}
  B.~A.~Dobrescu,
  Phys.\ Lett.\  B {\bf 461}, 99 (1999)
  [arXiv:hep-ph/9812349].



\bibitem{Cheng:1999bg}
  H.~C.~Cheng, B.~A.~Dobrescu and C.~T.~Hill,
  Nucl.\ Phys.\  B {\bf 589}, 249 (2000)
  [arXiv:hep-ph/9912343].


\bibitem{Davoudiasl:1999tf}
  H.~Davoudiasl, J.~L.~Hewett and T.~G.~Rizzo,
  Phys.\ Lett.\  B {\bf 473}, 43 (2000)
  [arXiv:hep-ph/9911262].
  
  
\bibitem{Chang:1999nh}
  S.~Chang, J.~Hisano, H.~Nakano, N.~Okada and M.~Yamaguchi,
  Phys.\ Rev.\  D {\bf 62}, 084025 (2000)
  [arXiv:hep-ph/9912498].


\bibitem{Rius:2001dd}
  N.~Rius and V.~Sanz,
  Phys.\ Rev.\  D {\bf 64}, 075006 (2001)
  [arXiv:hep-ph/0103086].
  


\bibitem{Gherghetta:2000qt}
  T.~Gherghetta and A.~Pomarol,
  Nucl.\ Phys.\  B {\bf 586}, 141 (2000)
  [arXiv:hep-ph/0003129].
  

\bibitem{Burdman:2007sx}
  G.~Burdman and L.~Da Rold,
  JHEP {\bf 0712}, 086 (2007)
  [arXiv:0710.0623 [hep-ph]].



  
\bibitem{Randall:1999ee}
  L.~Randall and R.~Sundrum,
  Phys.\ Rev.\ Lett.\  {\bf 83}, 3370 (1999)
  [arXiv:hep-ph/9905221].


\bibitem{Randall:1999vf}
  L.~Randall and R.~Sundrum,
  Phys.\ Rev.\ Lett.\  {\bf 83}, 4690 (1999)
  [arXiv:hep-th/9906064].


\bibitem{Uekusa:2010kh}
  N.~Uekusa,
  arXiv:1004.4410 [hep-ph].
  



\bibitem{Uekusa:2010jf}
  N.~Uekusa,
  arXiv:1002.1904 [hep-ph].



\bibitem{Burdman:2002se}
  G.~Burdman and Y.~Nomura,
  Nucl.\ Phys.\  B {\bf 656}, 3 (2003)
  [arXiv:hep-ph/0210257].



\bibitem{Uekusa:2008iz}
  N.~Uekusa,
  Int.\ J.\ Mod.\ Phys.\  A {\bf 23}, 3535 (2008)
  [arXiv:0803.1537 [hep-ph]].


\bibitem{Uekusa:2010ge}
  N.~Uekusa,
  Phys.\ Rev.\  D {\bf 82}, 065019 (2010)
  [arXiv:1006.5507 [hep-ph]].
  

\bibitem{Uekusa:2009pj}
  N.~Uekusa,
  arXiv:0908.0372 [hep-ph].


\bibitem{Raby:1979my}
  S.~Raby, S.~Dimopoulos and L.~Susskind,
  Nucl.\ Phys.\  B {\bf 169}, 373 (1980).

  
\end{thebibliography}
\end{document}